\documentclass[twocolumn]{aastex631}

\newcommand{\prot}{$P_{\mathrm{rot}}$}
\newcommand{\Msun}{\ifmmode {M_{\odot}}\else${M_{\odot}}$\fi}
\newcommand{\Rsun}{\ifmmode {R_{\odot}}\else${R_{\odot}}$\fi}
\newcommand{\Lsun}{\ifmmode {L_{\odot}}\else${L_{\odot}}$\fi}
\newcommand{\Rearth}{\ifmmode {R_{\oplus}}\else${R_{\oplus}}$\fi}
\newcommand{\lhal}{\ifmmode {L_{H\alpha}}\else$L_\mathrm{H\alpha}$\fi}

\newcommand{\lapprox}{{\lower0.8ex\hbox{$\buildrel <\over\sim$}}}
\newcommand{\gapprox}{{\lower0.8ex\hbox{$\buildrel >\over\sim$}}}

\def\asec{\ifmmode^{\prime\prime}\else$^{\prime\prime}$\fi}
\def\halpha{$\mathrm{H}\alpha$}
\def\LLH{$L_{\mathrm{H}\alpha}$/$L_{\mathrm{bol}}$}
\def\Lbol{$L_{\mathrm{bol}}$}

\def\prot{$P_{\mathrm{rot}}$}

\shorttitle{M Dwarfs in the TESS CVZ}
\shortauthors{Anthony et al.}
\graphicspath{{./}{figures/}}

\begin{document}

\title{\sc Activity and Rotation of Nearby Field M Dwarfs\\in the TESS Southern Continuous Viewing Zone}

\newcommand{\unc}{Department of Physics and Astronomy, University of North Carolina, Chapel Hill, NC 27599, USA}

\newcommand{\columbia}{Department of Astronomy, Columbia University, 550 West 120th St, New York, NY 10027, USA}

\newcommand{\lafayette}{Department of Physics, Lafayette College, 730 High St, Easton, PA 18042, USA}

\correspondingauthor{Francys Anthony}
\email{f.anthony@fisica.ufrn.br}

\author[0000-0003-1963-636X]{Francys Anthony}
\affiliation{Universidade Federal do Rio Grande do Norte (UFRN), Departamento de F\'isica, 59078-970, Natal, RN, Brazil} 

\author[0000-0002-8047-1982]{Alejandro N\'u{\~n}ez}
\altaffiliation{NSF MPS-Ascend Postdoctoral Research Fellow}
\affiliation{Department of Astronomy, Columbia University, 550 West 120th Street, New York, NY 10027, USA}

\author[0000-0001-7077-3664]{Marcel A.~Ag{\"u}eros}
\affiliation{Department of Astronomy, Columbia University, 550 West 120th Street, New York, NY 10027, USA}

\author[0000-0002-2792-134X]{Jason L.~Curtis}
\affiliation{Department of Astronomy, Columbia University, 550 West 120th Street, New York, NY 10027, USA}
\affiliation{Department of Astrophysics, American Museum of Natural History, 200 Central Park West, New York, NY 10024, US}

\author[0000-0001-7804-2145]{J.-D.~do~Nascimento,~Jr.}
\affiliation{Universidade Federal do Rio Grande do Norte (UFRN), Departamento de F\'isica, 59078-970, Natal, RN, Brazil}
\affiliation{Harvard-Smithsonian Center for Astrophysics, 60 Garden St., Cambridge, MA 02138, USA}

\author[0000-0002-5017-0628]{Jo\~ao M. Machado}
\affiliation{Universidade Federal do Rio Grande do Norte (UFRN), Departamento de F\'isica, 59078-970, Natal, RN, Brazil}

\author[0000-0003-3654-1602]{Andrew W.~Mann}
\affiliation{\unc}

\author[0000-0003-4150-841X]{Elisabeth R. Newton}
\affiliation{Department of Physics and Astronomy, Dartmouth College, Hanover, NH 03755, USA}

\author[0000-0001-7337-5936]{Rayna Rampalli}
\affiliation{Department of Physics and Astronomy, Dartmouth College, Hanover, NH 03755, USA}
\affiliation{Department of Astronomy, Columbia University, 550 West 120th Street, New York, NY 10027, USA}

\author[0000-0001-5729-6576]{Pa Chia Thao}
\altaffiliation{NSF Graduate Research Fellow}
\affiliation{\unc}

\author[0000-0001-7336-7725]{Mackenna L.~Wood}
\affiliation{\unc}

\begin{abstract}
{The evolution of magnetism in late-type dwarfs remains murky, as we can only weakly predict levels of activity for M dwarfs of a given mass and age. We report results from our spectroscopic survey of M dwarfs in the Southern Continuous Viewing Zone (CVZ) of the Transiting Exoplanet Survey Satellite (TESS). As the TESS CVZs overlap with those of the James Webb Space Telescope, our targets constitute a legacy sample for studies of nearby M dwarfs. For 122 stars, we obtained at least one $R\approx 2000$ optical spectrum with which we measure chromospheric \halpha\ emission, a proxy for magnetic field strength. The fraction of active stars is consistent with what is expected for field M dwarfs; as in previous studies, we find that late-type M dwarfs remain active for longer than their early type counterparts. While the TESS light curves for $\approx$20\% of our targets show modulations consistent with rotation, TESS systematics are not well enough understood for confident measurements of rotation periods (\prot) longer than half the length of an observing sector. We report periods for 12 stars for which we measure \prot~$\lapprox$~15 d or find confirmation for the TESS-derived \prot\ in the literature. Our sample of 21 \prot, which includes periods from the literature, is consistent with our targets being spun-down field stars. Finally, we examine the \halpha-to-bolometric luminosity distribution for our sample. Two stars are rotating fast enough to be magnetically saturated, but are not, hinting at the possibility that fast rotators may appear inactive in \halpha.}

\end{abstract}


\section{Introduction} \label{sec:intro}
M dwarfs are the most promising targets for searches of Earth-sized transiting exoplanets. Not only are the photometric signals for a given planet size larger than around e.g., a solar-like star, but transits of planets in the stellar habitable zone are more likely and more frequent \citep[][]{Gould2003}. Furthermore, based on results from the Kepler mission \citep{borucki2010}, such relatively close-orbiting planets are very common around M~dwarfs \citep{Dressing2015, Muirhead2015}. 

Striking examples of Earth-sized planets found orbiting late-type dwarfs by recent surveys include the seven orbiting the M8 star TRAPPIST-1 \citep{Gillon2017} and the 1.32~R$_{\oplus}$ planet transiting LHS 3844, another M~dwarf \citep{Vanderspek2019},   found in data from the first month of observations by the Transiting Exoplanet Survey Satellite \citep[TESS;][]{Ricker2015}. To constrain the evolution and habitability of these planets, however, we need to understand the magnetic activity of their parent stars today and in the past.

For example, \citet{garraffo2017} and \citet{Vida2017} found that flaring and winds from TRAPPIST-1{, by potentially stripping the atmospheres of its planets,} may have made these planets unsuitable for the development of life. {By contrast, \citet{Glazier2020} argued that the flare rate in TRAPPIST-1 is insufficient to catalyze chemical pathways thought to be required for ribonucleic acid synthesis, a key step in the chemical history of Earth. More generally, the flare duty cycle is an example of  stellar magnetic activity we need to understand, as it is thought to play a critical roll in the stability and/or depletion of Earth-like atmospheres around old M dwarfs \citep{Tilley2019, France2020}.
}

The activity--rotation relation is a powerful tool for exploring the magnetic behavior of late-type stars. In these stars, the \halpha~luminosity (\lhal), produced by magnetic heating of the chromosphere, increases as a fraction of its bolometric luminosity (\Lbol) with faster rotation. But this is true only up to a threshold velocity: stars rotating faster than this velocity show no further increase in activity levels \citep[][]{Noyes1984}, and are said to be magnetically saturated. Several theories have been advanced to explain this behavior, including changes in the magnetic dynamo \citep[e.g.,][]{Cameron1994} or in the topology of the magnetic field \citep[e.g.,][]{Garraffo2015}. Crucially, however, the exact mechanism driving the activity--rotation relation remains unknown, whether in the saturated or in the unsaturated regime.

The activity--rotation relation has been studied with samples encompassing a large range of stellar masses and rotational velocities  \citep[e.g.,][]{Delfosse1998, Jackson2010, Douglas2014, Nunez2015, Newton2017}. It remains challenging, however, to characterize this relation in the case of slower, less active rotators. Many studies focus on the stars in nearby, generally young ($\lapprox$1 Gyr) open clusters \citep[e.g., the $\approx$150-Myr-old NGC 2516 or the $\approx$700-Myr-old Hyades and Praesepe;][]{Jackson2010,Douglas2019,Rampalli2021}, where the periods are typically $\leq$20-25 d and the later-type stars tend to have strong \halpha\ emission. The field dwarfs we need to fill in our picture of the activity--rotation relation, in the regime that applies to planet hosts such as TRAPPIST-1, are presumably older and slower, with rotation periods that are more challenging to measure because of the long baseline photometric observations required. TESS data are an unparalleled opportunity to obtain \prot\ for these older, slow-rotating low-mass stars that are more representative of the overall Galactic population. TESS has a higher sensitivity to the redder spectra of later-type stars than its predecessor Kepler/K2 \citep{kepler2005, Ricker2015}. In addition, TESS spent its first year (2018 July--2019 July) monitoring the southern ecliptic pole Continuous Viewing Zone (CVZ) for 351~d, and has since then returned to observing the southern hemisphere. While it is currently challenging to confidently measure \prot\ longer than $\approx$15~d from TESS light curves alone, eventually it should be possible to use the resulting photometry, whether on its own or with that from other surveys \citep[e.g., the Zwicky Transient Facility in the north;][]{Bellm2019} to reliably measure very long periods for TESS targets. 

\begin{deluxetable}{@{}ll}
\tabletypesize{\small} 

\tablecaption{Properties of M Dwarfs in the Southern CVZ: Columns Headings \label{tbl:catalogcols}}

\tablehead{
\colhead{Column} & \colhead{Description} \\[-0.1 in]
}

\startdata
1      & 2MASS designation \\
2      & Gaia EDR3 designation \\
3, 4   & Epoch J2000 right ascension and declination\\
5, 6   & Distance and 1$\sigma$ uncertainty \\
7, 8   & $K$ band magnitude and 1$\sigma$ uncertainty \\
9, 10  & $r^\prime$ band magnitude and 1$\sigma$ uncertainty \\
11     & Source of $r^\prime$ magnitude\tablenotemark{a} \\
12, 13 & Gaia $G$ band magnitude and 1$\sigma$ uncertainty \\
14, 15 & Gaia $G_\mathrm{RP}$ band magnitude and 1$\sigma$ uncertainty \\
16     & Gaia RUWE \\ 17     & Number of spectra \\
18, 19 & EW and 1$\sigma$ uncertainty of the \halpha\ line \\
20     & Quiescent Absorption EW \\
21     & Rotation period \prot \\
22     & Source of \prot \tablenotemark{b} \\
23     & Stellar mass \\
24     & Convective turnover time $\tau$ \\
25,26 & Ratio $\chi$ of the continuum-to-bolometric flux\\
       & near the H$\alpha$ line, and 1$\sigma$ uncertainty\\
\enddata

\tablenotetext{a}{APASS: APASS DR9 Survey; DM09: Derived from the $V$--($J-K$)--$r^\prime$ relation in \citet{Dymock2009}.}
\tablenotetext{b}{AD17: Derived from the log($R^\prime_\mathrm{HK}$)--\prot\ relation in \citet{Astudillo2017}; Evryscope: The Evryscope \citep{Ratzloff2019}; ME: The MEarth Project \citep{Berta2012}; TESS: TESS.}
\end{deluxetable} 

We have conducted a spectroscopic survey of 125 M~dwarfs within 80 pc and located in the southern CVZ, obtaining at least one optical spectrum with which to measure the \halpha\ line strength for 122 of these stars. \citet{Gunning2014} found that the scatter in \halpha\ measurements for a group of field mid-M dwarfs was comparable to the full range of \halpha\ measurements exhibited by a single star. This implies that measuring \halpha\ only once may not capture the characteristic activity level of a single star, and that it is important to make multiple measurements to understand more fully the uncertainty on any individual measurement. We therefore obtained more than one spectrum for almost two-thirds of our stars, and 50\% have at least four spectra. {As the TESS CVZs overlap with e.g., the CVZs for the James Webb Space Telescope, our targets constitute a legacy sample for anyone interested in the properties of nearby field M dwarfs and/or of their exoplanets.}

In Section \ref{sec:data}, we describe the spectroscopic and photometric data for our sample of M dwarfs. In Section \ref{sec:properties}, we describe our \halpha\ and rotation-related measurements, from which we derive \LLH\ values and Rossby numbers for our stars. In Section \ref{sec:results}, we present our main results: \halpha\ equivalent-width measurements as a function of spectral type, and \LLH\ as a function of Rossby number, for our targets. In Section \ref{sec:conclusions}, we summarize our results and conclude.

\begin{figure}[!t]
    \centering
    \includegraphics[width=\columnwidth]{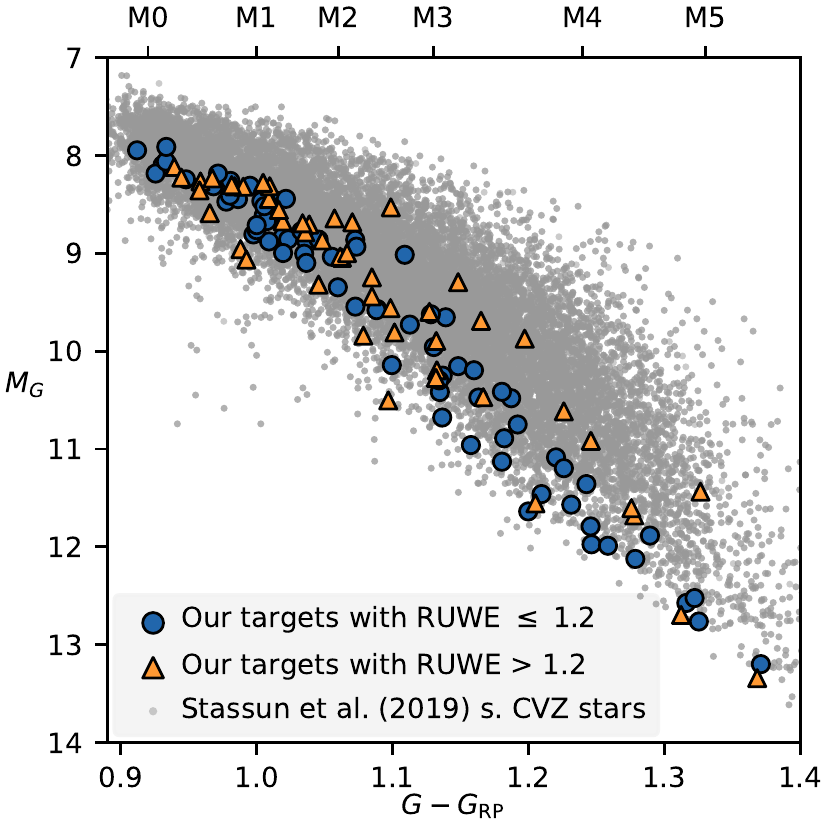}
    \caption{Gaia CMD showing our targets (blue circles: RUWE $\leq$~1.2; orange triangles: RUWE $>$~1.2) with, for comparison, 32,000 cool dwarfs in the TESS southern CVZ from the \citet{Stassun2019} catalog (gray dots).}
    \label{fig:cmd}
\end{figure}

\section{Sample Assembly, Spectroscopy, and Light Curves} \label{sec:data}

\subsection{Assembling our Sample}\label{sec:assembly}
We used the TESS input catalog Version 8 \citep[TICv8;][]{Stassun2019} to build our volume-limited sample of late-type dwarfs. First, we selected stars with masses $\leq$0.8~\Msun, within 100 pc, and located within 12$^\circ$ of the southern ecliptic pole. For this selection, we used the stellar masses and distances given by \cite{Stassun2019} in their specially curated Cool Dwarf list. {The lowest mass object selected in this manner has a mass of 0.13~\Msun.} Next, we searched for Gaia Early Data Release 3 \citep[EDR3;][]{Gaia2020} counterparts to stars in this sample. We then adopted the EDR3-based distances of \cite{Bailer-Jones2021} for the stars and, with these updated distances, recalculated their masses, using  
the \citet{Mann2019} mass--absolute {\it K} magnitude ($M_K$) relation \citep[as did][]{Stassun2019}, which is valid for $M_K$ between 4 and 10~mag. Finally, we selected all dwarfs from our initial sample with recalculated masses $\leq$0.6~\Msun\ and within 80 pc.

The resulting catalog contains 125 dwarfs with masses between 0.11 and 0.60 \Msun. Figure~\ref{fig:cmd} is a Gaia EDR3 color magnitude diagram (CMD) for our stars, with a selection of stars in the southern CVZ cataloged by \citet{Stassun2019} included for reference. 

\begin{figure}[!t]
    \centering
    \includegraphics{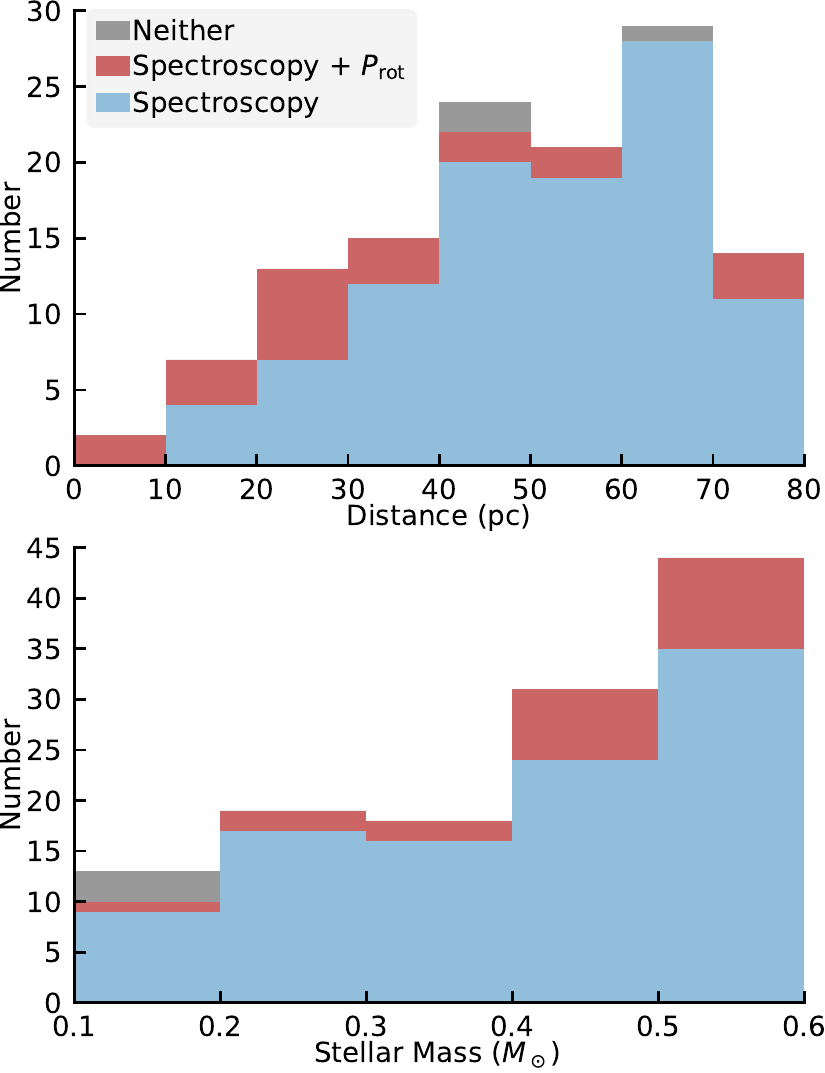}
    \caption{Our field dwarf sample, binned by distance (top panel) and stellar mass (bottom). The fraction of stars with spectra is shown in blue, that with both spectra and \prot\ measurements in red, and that  with neither in gray.}
    \label{fig:hist}
\end{figure}

Figure~\ref{fig:hist} gives the distance (top panel) and mass (bottom panel) distribution of our sample. Table \ref{tbl:catalogcols} describes the 26 columns in our catalog, {available in full online}. 

\cite{Stassun2019} did not attempt to identify potential binaries in the TICv8, and instead treated all objects as single stars when determining their properties. However, identifying these systems in our sample is essential, as members of binary and higher-order systems can have very different evolutionary histories that result in shorter \prot\ \citep[e.g.,][]{Meibom2007,Douglas2017} and potentially higher levels of \halpha\ activity than single stars of the same mass and age.

To help us identify potential binaries and higher-order systems, we used the Gaia EDR3 renormalized unit weight error (RUWE). The RUWE is a goodness-of-fit measure of the single-star model fit to the source's astrometry. When a star has a RUWE~$> 1.2$, there is a strong likelihood it has a companion, and the system is usually an unresolved binary (e.g., \citealt{Jorissen2019,Belokurov2020}). {The Gaia spatial resolving limit of $\approx$0\farcs7 \citep{Ziegler2018} corresponds to a semimajor axis $a\approx6$ au for the nearest candidate binary in our sample, and $\approx$54 au for the farthest. Stars in our sample with RUWE $> 1.2$ are indeed likely to be binaries with small enough separations for their components to have affected each other's protoplanetary disks in the first 10 Myr of their lifetimes \citep{Rebull2006, Meibom2007, Kraus2016, Messina2017}. This underlines the need to consider the observed properties of these stars separately from those of the likely single stars in our sample.\footnote{{In Figure~\ref{fig:cmd}, there are a handful of stars in our sample that appear brighter than stars of the same color, but that do not have RUWE $>$ 1.2. We believe that these are nonetheless likely binaries, as the RUWE does not capture all binaries, and our stars are on the whole older, field stars, and therefore unlikely to be e.g., young single stars that are overluminous due to their inflated radii.}}}
We include the RUWE for each star in our sample in Table~\ref{tbl:catalogcols}. Fifty stars (40\% of our sample) have RUWE $>$ 1.2, and we flag these in Figure~\ref{fig:cmd} and in our later analysis.

\begin{figure}
    \centering
    \includegraphics[width=.48\textwidth]{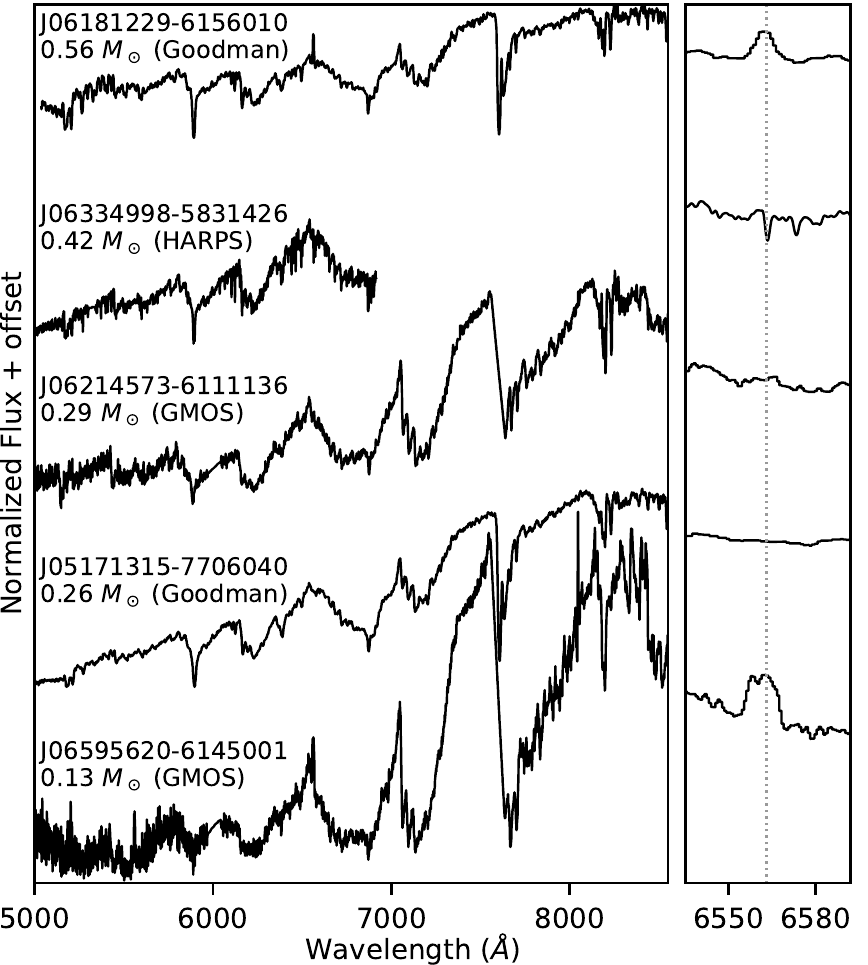}
\caption{Spectra of five of our M dwarfs ranging in mass from 0.13 to 0.56~\Msun. Four of these spectra were obtained using Goodman on SOAR or GMOS on Gemini-South; the fifth is an archival spectrum obtained with HARPS on the 3.6-m ESO telescope, La Silla. {\it Left---}Each spectrum is labeled with the star’s 2MASS designation and  mass. The spectra are normalized to the flux at 6555~\AA\ and smoothed to a resolution $R \approx 1000$. {\it Right---}A close-up of the \halpha\ line, the location of which is indicated with a vertical dotted line.}
    \label{fig:spec}
\end{figure}

\setlength{\tabcolsep}{2.5pt}
\begin{deluxetable}{lcc}
\tablewidth{0pt}
\tabletypesize{\small}
\tablecaption{SOAR and Gemini Observations \label{tbl:obs}}
\tablehead{
\colhead{UT Date} & \colhead{Instrument} & \colhead{\# of Spectra} 
}
\startdata
2019 Jan 10      & Goodman & 11  \\
2019 Feb 04      & Goodman & 29 \\ 
2019 Feb 16 - 17 & Goodman & 98  \\
2019 Feb 24      & Goodman & 20 \\
2019 Mar 08 - 09 & Goodman & 79 \\
2019 Apr 21      & Goodman & 27 \\ 
2019 May 10      & Goodman & 4 \\
2020 Jan 05 - 07 & Goodman & 92 \\
2020 Jan 14      & Goodman & 50 \\
2019 Jan - May      & GMOS & 54 \\
2019 Jul - 2020 Mar & GMOS & 26 \\
\hline
 & Total & 490
\enddata
\tablecomments{SOAR observations were obtained in part through programs LNA/SOAR SO2019A-014 (42 hours; 93\% completed) and SO2019B-015 (24 hours; 92\% completed), PI: do Nascimento. Gemini observations were obtained through Band 3 programs GS-2019A-Q-319 (20 hours; 100\% completed) and GS-2019B-Q-301 (17.18 hours; 42\% completed), PI: Ag\"ueros. }
\end{deluxetable} 
\subsection{Optical Spectroscopy}
We obtained our spectroscopic data  primarily through observations with the Goodman High-Throughput Spectrograph \citep[Goodman;][]{Clemens2004} on the Southern Astrophysical Research (SOAR) 4.1-m telescope and the Gemini Multi-Object Spectrograph \citep[GMOS;][]{gmoss} on the Gemini-South 8.1-m telescope, both located on Cerro Pach\'on, Chile. The bulk of our observations took place in early 2019, with additional data obtained at both observatories through early 2020 (see Table~\ref{tbl:obs}). We complemented these data with archival European Southern Observatory (ESO) spectra for eight of our targets.

Figure~\ref{fig:hist} shows our spectroscopic coverage. We have at least one spectrum for 122 of our targets, and a total of 551 spectra, which we use to characterize \halpha\ emission in these stars, as discussed in Section~~\ref{sec:halpha}. Five sample spectra for our stars are shown in Figure~\ref{fig:spec}. As can be seen in these examples, the spectra reveal a variety of \halpha\ emission levels in our target M dwarfs. 

\subsubsection{SOAR and Gemini Data}
At SOAR, we used the red camera with 400 $\ell/mm$ grating with the GG455 long pass filter, 1$\times$2 binning, and the 1.5\asec\ long slit. This configuration provided a wavelength coverage over 5000--9050 \AA\ with a maximum resolving power of $R\approx$~1850. We processed the spectra with the Goodman data-reduction pipeline \citep{GoodmanPipeline}\footnote{\url{https://github.com/soar-telescope/goodman_pipeline}}, which is designed to extract and wavelength-calibrate the spectra before doing the cosmic-ray removal and flux calibration. The resulting spectra have a signal-to-noise ratio (SNR) per pixel between 10 and 1000 at 6563~\AA. We have a total of 410 Goodman spectra for 80 stars.

At Gemini-South, we used the 400 $\ell/mm$ grating with the GG455 long pass filter, 1$\times$2 binning, and a 1.5\asec\ inner slit. This configuration gave us a wavelength coverage over 4500--9000 \AA\ with $R \approx$ 2000. We processed the spectra using the Gemini IRAF package v1.14\footnote{\url{https://www.gemini.edu/observing/phase-iii/understanding-and-processing-data/data-processing-software/gemini-iraf-general}} and followed the GMOS Data Reduction Cookbook\footnote{\url{http://ast.noao.edu/sites/default/files/GMOS_Cookbook/Processing/IrafProcLS.html}} to reduce long slit spectra. The resulting spectra have a SNR between 10 and 500 at 6563~\AA. We have a total of 80 GMOS spectra for 78 stars. Thirty-six stars have both Goodman and GMOS spectra.

\subsubsection{Archival ESO Data}\label{sec:ESO}
We retrieved 61 reduced spectra for eight stars in our sample from the ESO Science Archive.\footnote{These spectra were processed using the ESO calibration pipeline v3.5 or v3.8: \url{http://www.eso.org/sci/observing/phase3.html}} These spectra were obtained with the High Accuracy Radial velocity Planet Searcher \citep[HARPS;][]{Mayor2003}, a high-resolution echelle spectrograph, on the 3.6-m ESO telescope, La Silla, Chile, between 2003 December and 2020 March as part of eight different scientific programs. For all of these stars, we also have a Goodman or GMOS spectrum, so that we have a total of 551 spectra for 122 stars.  
The spectra have a wavelength coverage over 3800--6900~\AA\ with $R \approx$ 115,000 and a SNR between 16 and 120 at 6563~\AA.

\subsection{Optical Light Curves} \label{sec:lightcurves}

We extracted TESS photometric data using publicly released full-frame images (FFIs) to create 30-minute cadence light curves for all 125 stars in our sample. First, we downloaded a $21 \times 21$ pixel cutout image centered on each target via the \texttt{TESScut} tool on the Mikulski Archive for Space Telescopes server \citep{Brasseur2019}. 
We used a circular aperture of radius 2 pixels around the object to perform the photometric extraction. We estimated the background signal to be subtracted from the light curve through a sigma ($\sigma$) clip over the entire FFI cutout using the \texttt{photutils} algorithm \citep{Bradley2016}.
The flux for each TESS sector (corresponding to a 27.4~d observing window) was divided by the median to normalize and, consequently, preserve the light curve variability. We performed all these steps using the \texttt{tesseract} pipeline\footnote{ \url{https://github.com/astrofelipe/tesseract}}. Finally, we stitched together data for all the available sectors from the first year of TESS operations for each of our targets. 

The resulting light curves were carefully processed to remove data affected by systematic effects that clearly impacted their quality. For a fraction of the stars, we identified abrupt flux increases or decreases at the beginning and/or end of some sectors, such as the side the gaps created when the data were being sent to Earth. We attempted to mitigate these effects by discarding photometry offset from the median by more than 5$\sigma$. Even after this 5$\sigma$-clipping, however, there were still obvious systematics present in the data. {Because of these issues, we also used the Causal Pixel Modeling approach \citep[CPM;][]{Hattori2021} to re-extract the light curves and confirm the results reported in Section~\ref{sec:prot}.} 

\begin{figure*}[t]
    \centering
    \includegraphics[scale=1]{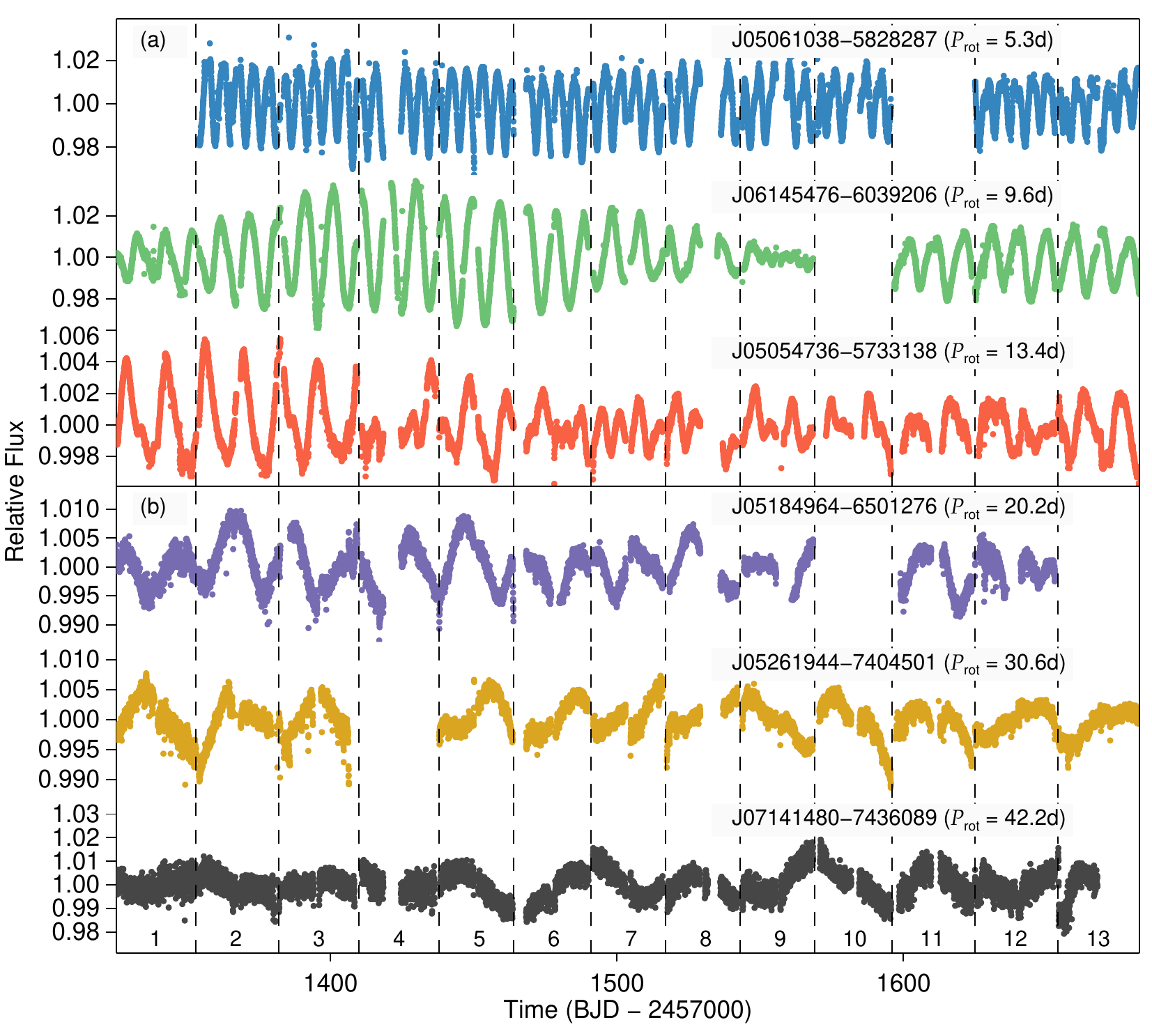}
    \caption{TESS light curves for M dwarfs in our sample observed during the first year of its mission. Dashed vertical lines indicate the start and end of each observing sector. The number of each sector is labeled at the bottom. Each light curve is labeled with the star’s 2MASS designation and the \prot\ obtained from our GLS analysis.  Panel (a) shows examples of stars with \prot\ $<$15 d, which we considered our upper limit for confident \prot\ measurements based exclusively on TESS data. Panel (b) shows examples of light curves where we measured \prot~$>$~15~d, and thus considered unreliable. The first two stars in Panel (b) have \prot\ measurements in \citet{Howard2020} that disagree with those measured here by more than 20\%. As for the third star, while its light curve shows plausible modulations on a timescale of $\approx$40 d, we lack any ancillary data to confirm the TESS-derived \prot. 
}
    \label{fig:lc}
\end{figure*}

We also examined the apertures to check whether there was contamination from nearby bright sources, which would impact our ability to assign confidently the observed light curve behavior of our target. 
In 20 cases, we found a source near our target that had a magnitude difference of less than 1 mag, and we removed those from the analysis. 

About one-fifth (i.e., 24/125) of the stars in our sample exhibit modulations consistent with rotation. Figure \ref{fig:lc} shows  examples of these light curves. As discussed in more detail in Section~\ref{sec:prot}, our efforts did not necessarily produce light curves of sufficient quality to extract credible \prot\ values longer than about half the length of a TESS {sector} length, or $\approx$15~d. This is mainly due to uncertainties introduced in the processing necessary to stitch together the light curves from the different sectors, which may either remove or introduce photometric variations.

\section{Rotational Properties and \halpha\ measurements of Our Targets}\label{sec:properties}

\subsection{Measuring Rotation Periods}\label{sec:prot}

We used the generalized Lomb--Scargle \citep[GLS;][]{Zechmeister2009} periodogram to measure \prot\ from our TESS light curves. The GLS periodogram takes into account measurement errors when finding the best fit to the data, and generally avoids aliasing in the period determination, and can therefore be considered superior to the standard Lomb-Scargle approach \citep{VanderPlas2018,Reinhold2020}. We used the false alarm probability (FAP) to assess the significance of the power of the periodogram peaks, considering only \prot\ for periodograms with a FAP $<$ 1\%. We obtained \prot\ measurements for 24 {out of 125 stars in our intial sample, which represents $\approx$20\% of the total.}

However, as was highlighted recently by \cite{Claytor2021}, there are fundamental issues with trying to measure longer \prot\ with TESS light curves produced by stitching together data from several 27-d sectors. In the resulting light curves, it is difficult to assess whether a signal is caused by true fluctuations in stellar brightness, or whether it is a result of systematics in the data either incorrectly removed or added in the processing of the light curves. It was noticeable, and disconcerting, that we found a number of stars for which the apparent flux modulations (and resulting \prot\ measurements) coincided with the length of a sector. 

As a result, our position is that confidently measuring periods for slow rotators (defined here as having \prot~$\gapprox$~15~d) based only on TESS light curves is challenging, and that reporting these \prot\ is potentially misleading. We therefore initially limited our TESS-derived sample of \prot\ to those stars for which \prot~$\lapprox$~15 d. {With this restriction, our final sample of reliable \prot~from TESS light curves is 12 stars. For each of these objects, we include in the two figure sets in the Appendix the $21\times21$ pixel cutout of the TESS FFI used to extract the light curve, the extracted light curve, the GLS periodogram, and the phase-folded light curve.}

Given these limitations in TESS data, we searched the literature for other measurements of \prot\ for stars in our sample. We found six slow-rotating stars whose periods were inferred by \citet{Astudillo2017} by applying an empirical relation between the strength of the Ca II H and K emission lines, another tracer of magnetic activity, and the \prot\ of M dwarfs. Additionally, we found one rotator in the MEarth Project \citep{Berta2012} database\footnote{\url{https://lweb.cfa.harvard.edu/MEarth/DR10/}} that is also in our sample, and we used the MEarth \prot\ to confirm our TESS-derived period of {14.34} d for that star. Finally, we found two stars in our sample with long rotation periods extracted from Evryscope \citep{Ratzloff2019} light curves by \citet{Howard2020}. These two stars are the ones shown in Panel (b) of Figure \ref{fig:lc}, and we adopted the Evryscope \prot\ for them. In this fashion, we increased the size of our sample of rotators to 21 stars.

\subsection{Measuring H{\ensuremath{\alpha}} EWs and Estimating Quiescent EWs}\label{sec:halpha}

We used the code \texttt{PHEW} \citep{Alam2016}, which uses \texttt{PySpecKit} \citep{Ginsburg2011} to fit lines with a Voigt profile, to measure automatically the equivalent width (EW) of the \halpha\ line, and to estimate the corresponding EW uncertainty, 
for all of our spectra.

We interactively defined continuum regions blueward and redward of the \halpha\ line in each spectrum. These continuum regions varied in length between 5 and 35~\AA\ on each side, but the total length of the continuum considered was never less than 20 \AA.

We extracted the noise for each point in our spectra from a Gaussian width equal to the $\sigma$ of the flux in the continuum regions defined for the individual spectra.
\texttt{PHEW} performed 1000 Monte Carlo iterations adding this noise to the spectrum and calculated the mean and $\sigma$ of the 1000 resulting EWs, which we assigned as the EW value and 1$\sigma$ uncertainty for that spectrum. When a star had multiple spectra, we took the error-weighted mean EW and the weighted mean standard error to be the EW value and 1$\sigma$ uncertainty for that star. The overwhelming majority of the stars in our sample with multiple spectra showed little evidence of variability in the \halpha\ EW we measured from their spectra.

We measured an \halpha\ EW value for 133 of the 551 spectra in our sample. For the other 438, the \halpha\ line was indistinguishable from zero. 
Our EW values range between $-11.5$ and 1.1 \AA. 
The smallest value we could measure for a star with \halpha\ emission was EW $=-0.5$ \AA. 
For the spectra with \halpha\ indistinguishable from zero, we assigned EW = 0 \AA\ and a systematic EW 1$\sigma$ uncertainty of 0.5 \AA.

The eight stars in our sample with high-resolution HARPS spectra also have low-resolution (GMOS or SOAR) spectra. In six of these stars, the low-resolution spectra exhibited an \halpha\ line indistinguishable from zero, whereas the high-resolution spectra showed a measurable \halpha\ line in absorption, covering the EW range 0.2-0.5 \AA. {The lack of \halpha\ absorption in the GMOS/SOAR spectra for these six stars reflects the fact that below $\approx$0.5 \AA, we reach the smallest resolvable wavelength difference in these spectra. Still,} in the other two stars, both low- and high-resolution spectra showed \halpha\ in absorption covering the range 0.3-0.5 \AA, {although the low-resolution spectra have large ($>$20\%) EW uncertainties.}

\begin{figure*}
    \centerline{\includegraphics[scale=1.35]{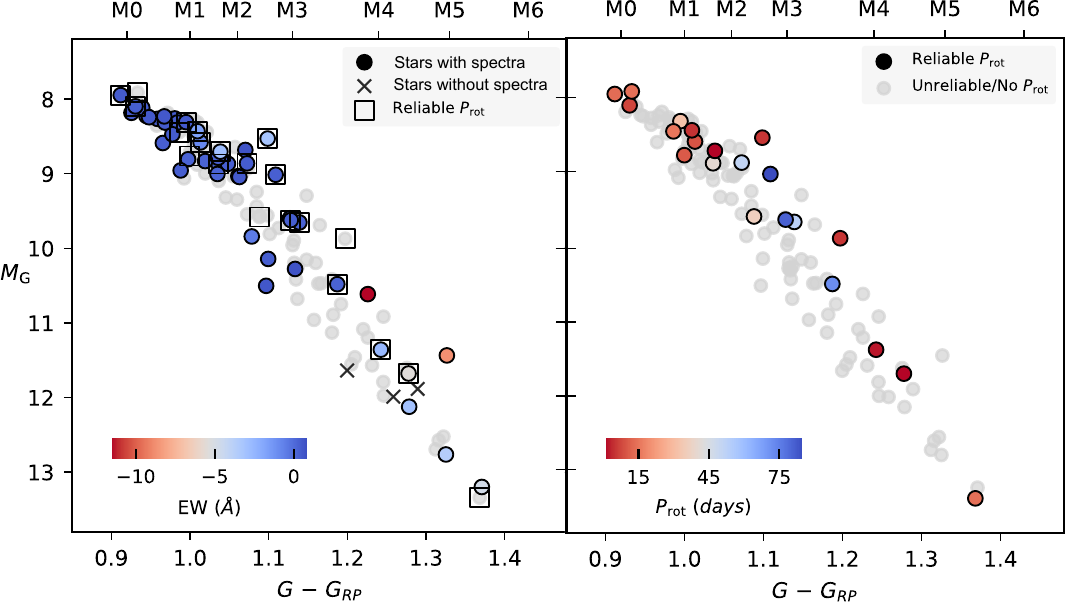}}
    \caption{Gaia CMDs showing the distribution of stars with EW and \prot\ measurements in our sample. {\it Left---}The circles are color-coded according to the EW value using the colorbar at the bottom left. Gray circles are stars to which we assigned
    EW = 0 (see Section~\ref{sec:halpha}). Stars without spectroscopy are indicated with $\times$ symbols. The open squares indicate stars with reliable \prot\ (see Section~\ref{sec:prot}). {\it Right---}The circles are color-coded according to the \prot\ value using the colorbar at the bottom left. Gray circles are stars in our sample with no \prot\ measurement or unreliable determination.}
    \label{fig:CMDew}
\end{figure*}

All in all, we have 45 stars (37\% of our sample) with at least one spectrum with a non-zero EW and 77 stars (62\% of our sample) with EW = 0~\AA. In the left panel of Figure~\ref{fig:CMDew}, we color-code the stars in our sample on their Gaia CMD by the strength of their \halpha~EW measurement. The three stars for which we have no spectra are indicated with $\times$ symbols, and stars with trustworthy \prot\ measurements are highlighted with open squares (see Section~\ref{sec:prot}). {In turn, the right panel shows the \prot\ distribution of these 21 stars on a Gaia CMD.} 

\citet{Stauffer1986} showed that in inactive M dwarfs, the \halpha\ line appears in absorption;  the line fills in and eventually transitions to emission as the stars become more and more active. Accurately measuring the EW of the \halpha\ feature therefore requires considering the basal absorption level, which is a function of spectral type. We calculated this quiescent absorption EW value for each of our stars using the empirical model of \citet{Newton2017} and then determined the relative EW, the measured EW minus the quiescent absorption EW. Table~\ref{tbl:catalogcols} includes both the measured EW (also plotted in Figure~\ref{fig:CMDew}) and the calculated quiescent absorption EW for the 122 stars in our sample with spectroscopy.

\begin{figure}[!t]
    \centerline{\includegraphics{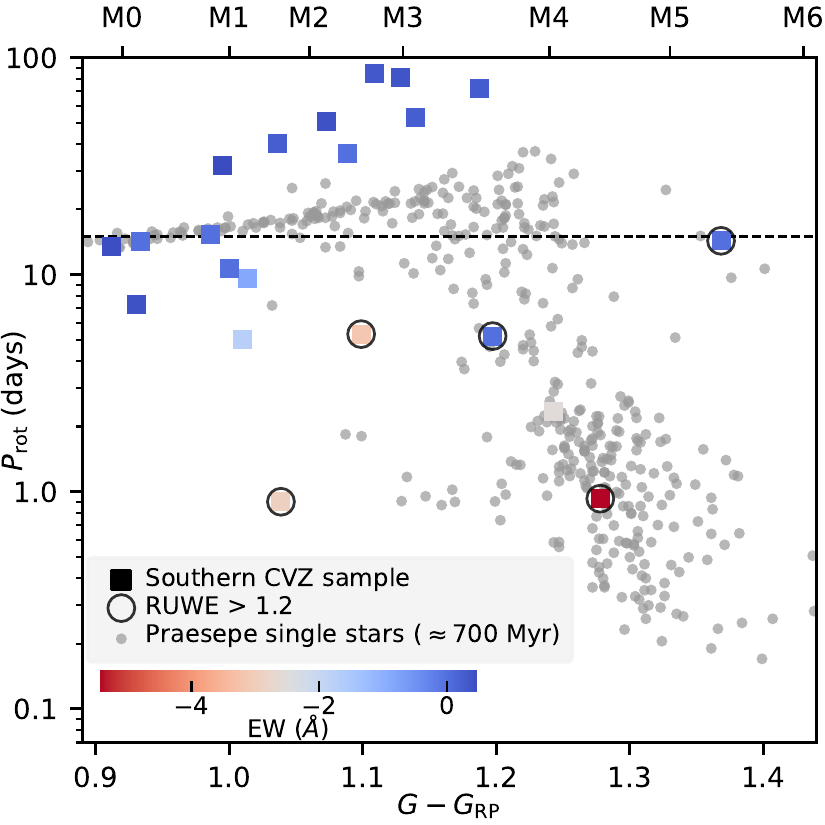}}
    \caption{\prot\ vs.~Gaia ($G-G_\mathrm{RP}$) color for stars in our sample, represented with square symbols. Black circles edges indicate stars with RUWE $>$ 1.2. The distribution for Praesepe single stars 
    \citep[$\approx$700 Myr old; from][]{Rampalli2021} is plotted in gray in the background. The squares are color-coded according to the \halpha\ EW value using the colorbar at the bottom left. The horizontal dash line indicates our upper limit in \prot\ space when calculating \prot\ from TESS light curves (see Section~\ref{sec:prot}).}
    \label{fig:Prots}
\end{figure}

The color--period distribution for these 21 stars is shown in Figure~\ref{fig:Prots}. We also indicate which of these rotators have  measured \halpha\ EWs. Comparing our sample of field rotators to rotators in the $\approx$700 Myr-old Praesepe open cluster (gray dots in Figure~\ref{fig:Prots}) underlines the difference in the ages of the two samples: the majority of the likely single stars in our sample have larger \prot\ values than the stars that define the slow rotating sequence in Praesepe, which extends from K7 to M3.5 and \prot~$\approx$~15 to 30 d. This is consistent with our sample being populated by older, spun-down field stars. 

\subsection{Obtaining \LLH}\label{sec:LLH}
\LLH\ is a measure of the fractional contribution of \halpha\ emission relative to the entire energy output of the star, and is therefore a more useful indicator of chromospheric activity than \lhal\ when comparing stars spanning a range of masses. 

We obtained \LLH\ for stars with \halpha\ in emission by using the relation
\begin{equation}
    L_{\mathrm{H}\alpha}/L_{\mathrm{bol}} = -W_{\mathrm{H}\alpha} \frac{f_0}{f_{\mathrm{bol}}},
\label{eq:LLH}
\end{equation}
where $W_{\mathrm{H}_{\alpha}}$ is the relative \halpha\ EW, $f_0$ is the continuum flux near the \halpha\ line, and $f_{\mathrm{bol}}$ is the apparent bolometric flux. Empirically, one can calculate $f_0 / f_{\mathrm{bol}}$, known as $\chi$, as a function of color. \citet{Douglas2014} derived a relation between $\chi$ and color using spectra from the \textsc{Phoenix Aces} model spectra \citep{Husser2013} with solar metallicity, log($g$) = 5.0, and 2500 $\leq T_{\mathrm{eff}} \leq$ 5200~K. 

We used the log($\chi$)--($r^\prime-K$) relation in \citet{Douglas2014} to derive $\chi$ values for our stars. We obtained $r^\prime$ photometry from the AAVSO Photometric All Sky Survey \citep[APASS;][]{Henden2016}, and $K$ photometry from the Two Micron All Sky Survey \citep[2MASS;][]{2mass}. For the 20 stars in our sample with no APASS photometry, we used $V$ photometry from the TICv8 catalog and the $V$--($J-K$)--$r^\prime$ relation from \citet{Dymock2009} to estimate $r^\prime$. The $K$ and $r^\prime$ photometry we used for each star is included in Table~\ref{tbl:catalogcols}.

We did not calculate \LLH\ for stars with EW uncertainties large enough to cross over into potential \halpha\ absorption. This was the case for all stars for which we assigned an EW = 0 \AA.

\subsection{Calculating Rossby Numbers}
The scatter in the activity--rotation relationship is significantly smaller if rotation is parameterized in the form of the Rossby number, defined as $R_\mathrm{o} = P_{\mathrm{rot}} / \tau$, where $\tau$ is the convective turnover time \citep[e.g.,][]{Noyes1984, Wright2011}. We estimated $\tau$ values for our stars using the empirical mass--log($\tau$) relation of \citet{Wright2018}, which is based on \prot\ and X-ray luminosity measurements for almost 850 stars in the mass range 0.08--1.36~\Msun. With those $\tau$ values, we calculated $R_\mathrm{o}$ for the 21 stars in our sample with a measured \prot. Table~\ref{tbl:catalogcols} includes the $\tau$ values we calculated for our stars.

\section{Results and Discussion} \label{sec:results}

\subsection{H{\ensuremath{\alpha}} EW Distribution}
\begin{figure}
    \centering
    \includegraphics[width=.47\textwidth]{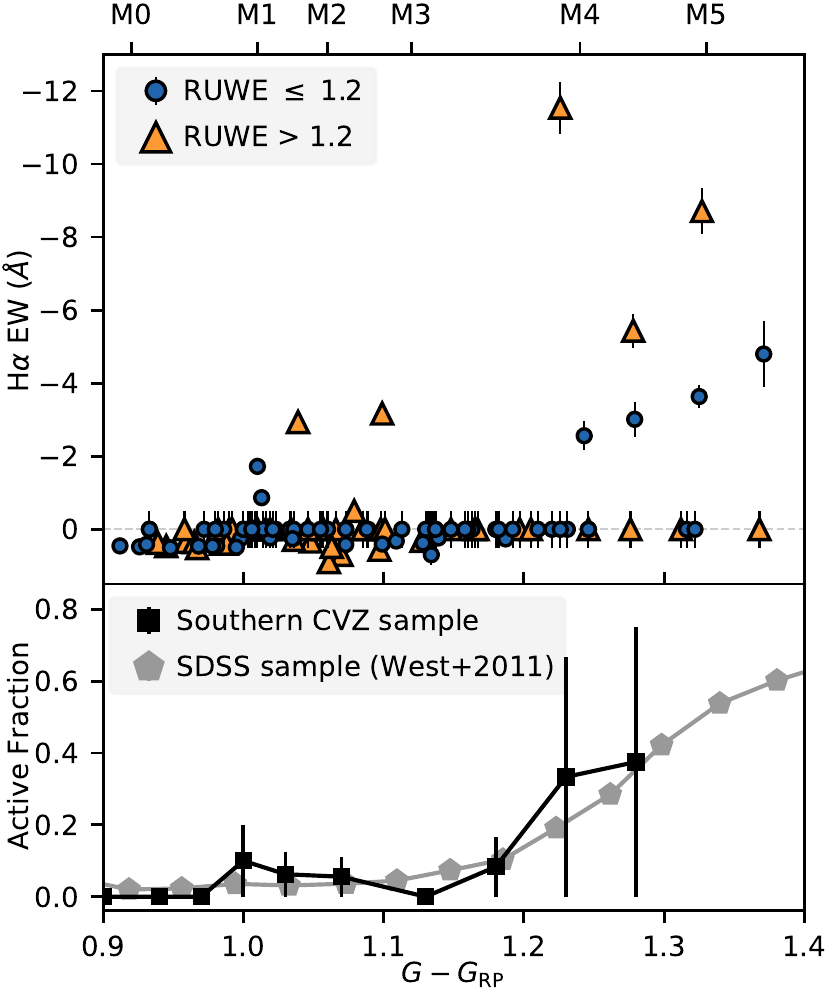}
    \caption{{\it Top---}\halpha\ EW as a function of Gaia ($G-G_\mathrm{RP}$) color and spectral type for stars in our sample. Blue circles indicate stars with Gaia's RUWE $\leq$ 1.2, and orange triangles, stars with RUWE $>$ 1.2; the latter are potential unresolved binary or higher-order systems. {\it Bottom---}A comparison of the active fraction of M dwarfs in our sample (black squares and line) to that for the sample of M dwarfs with SDSS spectra of \citet[][grey pentagons and line]{West2011}. We define active/inactive stars following the \halpha\ EW--($G - G_\mathrm{RP}$) boundary relation in \citet{Kiman2021}.}
    \label{fig:eqw}
\end{figure}

The top panel of Figure~\ref{fig:eqw} shows the \halpha\ EW as function of Gaia color and spectral type for the 122 stars in our sample with spectroscopy. We used the \cite{Kiman2021} definition for the \halpha\ EW boundary as a function of ($G - G_\mathrm{RP}$) between active and inactive stars. Stars earlier than M3 appear largely to be inactive, while a larger portion of stars later than that spectral type---corresponding to the transition between Sun-like and fully convective stars {\citep{Chabrier1997}}---are active. This behavior is consistent with the assumption that a sample of field stars should generally be older than a few Gyr. Analyses of \halpha\ activity in younger single-aged populations such as Praesepe and the Hyades find that the early M dwarfs are active \citep[see, e.g., figure 8 of][]{Douglas2014} but, by $\approx$1~Gyr, they no longer show evidence for \halpha\ emission  \citep[e.g., the M0-M2 stars in the $\approx$1.3-Gyr-old open cluster NGC 752; see figure 11 of][]{agueros2018}.

In Figure~\ref{fig:eqw}, we indicate with orange triangles stars in our sample with RUWE $>$ 1.2, and thus suspected of being binaries. While there does not seem to be a difference between the overall behavior of potential binaries compared to single stars in this plot, we do note more elevated levels of \halpha\ emission in some of the later-type potential binaries. {Typically, magnetic activity, and hence \halpha\ emission, can be increased by the presence of accretion disks \citep[e.g.,][]{Muzerolle1998, Biazzo2012} or by interactions between members of close binaries \citep{Morgan2012, Skinner2017}. Although accretion disks have long disappeared at the expected ages of the field stars in our sample \citep[e.g.,][]{Haisch2001, Lee2020}, the higher \halpha\ levels for stars with high RUWE may be interpreted as evidence of enhanced magnetic activity due to a nearby companion. As mentioned in Section~\ref{sec:assembly}, the stars in our sample with high RUWE are likely interacting binaries, potentially resulting in increased activity levels relative to single stars of the same mass. However, a binary identification based on RUWE alone would miss tight ($a~\lapprox~0.1$~au), tidally interacting binaries \citep[e.g.,][]{Belokurov2020}. And the enhanced activity in our stars with high RUWE could be driven instead (or additionally) by the strong magnetic interactions in such compact stellar systems. Further observations are required to understand whether stars in our sample with high RUWE have additional, hidden companions.}

To provide more context for these data, the bottom panel of Figure~\ref{fig:eqw} shows the active fraction of stars in our sample as a function of the binned Gaia color (black squares). As in \citet{Kiman2021}, we calculated uncertainties for each bin, based on a binomial distribution, as $\sigma_f = (f \times (1 - f))/n$, where $f$ is the active fraction and $n$ is the bin size.
We compared the active fraction of our sample with that of {more than $7\times10^4$} Sloan Digital Sky Survey \citep[SDSS;][]{york00} M dwarfs from \citet[][gray pentagons in Figure~\ref{fig:eqw}]{West2011}.\footnote{\cite{Kiman2021} did not correct their \halpha\ data to account for the basal absorption in inactive stars. We therefore used our measured EW values for this comparison instead of our calculated relative EW values.} The fraction of active stars in our sample is consistent with what is expected for field {(i.e., $\gapprox$1 Gyr)} M dwarfs: close to zero for early M dwarfs, and $\approx$0.4 for mid-M dwarfs. {As in previous studies of M dwarfs \citep{West2004, West2011, Schmidt2015, Kiman2021}, we conclude that later type M dwarfs remain active for longer than their earlier brethren.}

\subsection{Activity--Rotation Relation}

\begin{figure}
    \centering
    \vspace{.2cm}
    \includegraphics[width=0.47\textwidth]{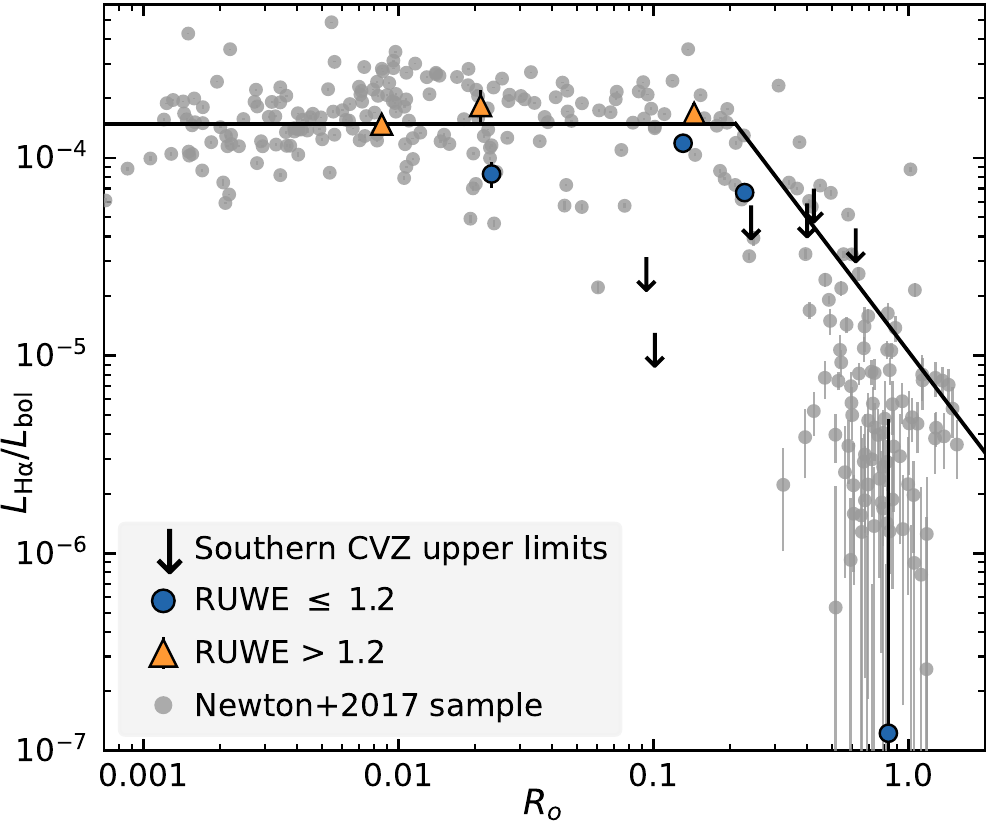}
    \caption{\LLH\ as a function of $R_\mathrm{o}$ for our southern CVZ sample and for the M dwarfs of \citet[][gray circles]{Newton2017}. Blue circles indicate stars in our sample with RUWE $\leq$ 1.2, and orange triangles, with RUWE $>$ 1.2. The arrows indicate upper limits for stars in our sample with \halpha\ indistinguishable from zero; for these stars we assumed an EW upper limit of $-0.5$ \AA\ to measure their \LLH\ upper limit. The black line is the best fit from \cite{Newton2017}.}
    \label{fig:Lalpha}
\end{figure}

In Figure~\ref{fig:Lalpha}, we plot \LLH\ as a function of $R_\mathrm{o}$ for the stars in our sample with a \LLH\ measurement (see Section~\ref{sec:LLH}). For context, we also show the {large sample of nearby M dwarfs} of \citet{Newton2017}. {These authors' sample contains more than 400 stars with \prot\ values, nearly all of which were measured from MEarth light curves, as well as \halpha\ relative EW values, i.e., corrected for quiescent \halpha\ absorption (see Section~\ref{sec:halpha}). The \citet{Newton2017} sample} exhibits {a typical} saturated/unsaturated behavior {\citep[e.g.,][]{Mohanty2003, Reiners2012, Douglas2014}}: stars with $R_\mathrm{o}\ \gapprox\ 0.2$ show a power-law dependence of \LLH\ on rotation, while stars with a $R_\mathrm{o}$ below this threshold are saturated, with \LLH~$\approx$~constant and independent of rotation. {Our stars follow the same distribution as the \citet{Newton2017} sample.}

We also include \LLH\ upper limits for stars with \halpha\ indistinguishable from zero in our spectra; for these stars, we assumed an EW upper limit of $-0.5$ \AA\ to estimate their \LLH\ upper limit. {As can be seen in Figure~\ref{fig:Lalpha}, two of our \LLH\ upper limits fall almost one order of magnitude below the typical value for their $R_\mathrm{o} \approx 0.1$. These M dwarfs are rotating fast enough that they should be magnetically saturated, and yet they are not. While \citet{Newton2017} showed clearly that M dwarfs without detectable \halpha\ emission are slow rotators, the contrapositive remained unanswered: are all fast rotators \halpha\ emitters?}

{The answer appears to be no, although the number of examples of fast-rotating M dwarfs that do not show \halpha\ emission is small. \citet{Newton2017} found one fast-rotating, \halpha-inactive star in their sample. Earlier, \citet{West2009} identified three fast rotators in a sample of 14 M6-M7 dwarfs selected because of their unusually low levels of \halpha\ emission in SDSS spectra. There are therefore intriguing suggestions that some fast-rotating M dwarfs may appear inactive in \halpha. Follow-up X-ray observations of these stars to measure their level of coronal activity, perhaps with the eROSITA instrument aboard the Spectrum-Roentgen-Gamma mission \citep{Sunyaev2021}, could help to determine whether these unusual stars are truly magnetically inactive.}

{Lastly, we see in Figure~\ref{fig:Lalpha} no distinct behavior for stars in our sample with RUWE $>$ 1.2. If these stars are indeed unresolved binaries, the fact that their \LLH\ are comparable to the \LLH\ of stars with similar $R_\mathrm{o}$ in our sample and in the \citet{Newton2017} sample indicates that their \LLH\ measurements are likely dominated by the \halpha\ emission of only one of the binary components.}

Our ability to contribute data to this plot is currently limited by the small number of stars for which we can confidently measure \prot\ with TESS data. These are primarily fast rotators that appear saturated in this plot. {However, there are reasons to be optimistic about our ability to overcome the systematic problems arising from stitching TESS sectors. One promising approach has been to combine ground and space data \citep[e.g.,][]{Andrews2021, Howard2021}.} With the hoped-for improvements in our confidence in measuring longer \prot\ from TESS light curves (or by combining TESS light curves with data from other surveys), we aim to eventually add more points to the sample of unsaturated stars shown in Figure~\ref{fig:Lalpha}.

\section{Conclusions} \label{sec:conclusions}
We have conducted a spectroscopic survey of M dwarfs located in the southern TESS CVZ and within 80 pc of the Sun. We used the TICv8 \citep[][]{Stassun2019} to build our volume-limited sample of 125 stars, and, through a combination of observations with SOAR/Goodman and Gemini-South/GMOS and archival ESO/HARPS data, obtained at least one optical spectrum with which to measure the \halpha\ line strength for 122 of these stars.

Because measuring \halpha\ only once may not capture the characteristic activity level of a single star \citep{Gunning2014}, it is important to make multiple measurements to understand more fully the uncertainty on any individual measurement. We therefore obtained more than one spectrum for almost two thirds of our targets. These reduced spectra are publicly available from the Columbia University Academic Commons, an online research repository\footnote{Available on the Columbia Academic Commons: \dataset[doi:10.7916/d8-xg58-9x44]{https://doi.org/10.7916/d8-xg58-9x44}.}.

We measured the \halpha\ EW for all of the spectra in our sample. For stars with multiple spectra, we took the error-weighted mean EW and the weighted mean standard error as the representative EW and uncertainty. For the majority of our stars, we found an \halpha\ line strength indistinguishable from zero, and in the cases where we had more than {one} spectra, no evidence for variability in the measured line strength.

We calculated quiescent absorption EW values for our stars using the \citet{Newton2017} model and then determined the relative EW for each star by combining the quiescent and the measured EW values. Stars earlier than M3 in our sample appear largely inactive, while a larger fraction of stars later than that spectral type are active. This behavior is consistent with the assumption that a sample of field stars should be older than a few~Gyr. We also compared the active fraction of our sample with that of a large sample of M dwarfs with SDSS spectra \citep{West2011}. The fraction of active stars in our sample is consistent with what is expected for field M dwarfs.

We generated photometric light curves for all of our targets using the publicly released TESS FFIs. For each star, the flux for each TESS sector was divided by the median value for that sector, and we then stitched together data for all the available sectors. The resulting, nearly year-long light curves were processed to remove data affected by systematics. This processing, however, is still far from perfect, with TESS systematics not yet well enough understood for confident measurement of \prot\ longer than roughly half the length of a sector, or $\approx$15 d \citep{Claytor2021}. 

While about one-fifth of the stars in our sample exhibit modulations consistent with rotation, we report TESS periods only for those stars for which we either measure \prot~$\lapprox$~15 d with TESS data alone, or for which we find confirmation for the TESS-derived \prot\ in the literature. There are 12 stars in the former category, and another nine whose periods we confirm or obtain using the literature. Comparing our sample to rotators in the $\approx$700 Myr-old Praesepe cluster underlines the difference in the ages of the two samples: the majority of the likely single stars in our sample have longer \prot\ than the stars that define the slow rotating sequence for K7 to M3.5 stars in Praesepe. This is consistent with our sample containing older, spun-down field stars.

We estimated \LLH\ and $R_\mathrm{o}$ for our stars and compared these to the distribution of field M dwarfs assembled by \citet{Newton2017}. Our ability to contribute data to this distribution is currently hindered by the small number of stars for which we can confidently measure \prot~with TESS data. These are a handful of fast and slow rotators that are either magnetically saturated, with \LLH~$\approx$~10$^{-4}$ for the stars with $R_\mathrm{o}$~$<$~0.2, or that exhibit an activity level consistent with what is expected given their larger $R_\mathrm{o}$. {Two stars in our sample with $R_\mathrm{o}$~$<$~0.2 only have \LLH\ upper limits. Together with a handful of other similar outliers in the literature, they hint at the possibility that fast rotators may appear inactive in \halpha. Thus, follow-up observations, e.g., of their coronal X-ray emission, are necessary to determine whether these are truly rapidly rotating, but magnetically inactive, M dwarfs.}

{The full value of TESS data will be realized when we learn to combine TESS sectors to obtain confident \prot\ measurements for the slowest rotators.} With the hoped-for improvements in our confidence in measuring longer \prot\ from TESS light curves, we aim  eventually to add more points to the sample of unsaturated stars. Our combination of \halpha\ and \prot\ measurements can potentially help constrain the activity-rotation relation for the slowest rotators, {as TESS will produce some of its longest-baseline light curves, spread over several years of its observing campaigns, for this sample of M dwarfs.
}

\begin{acknowledgments}
{We thank the anonymous referee for their careful reading of the manuscript and their insightful suggestions.}

We thank S.~T.~Douglas for assistance with the spectroscopic observing proposals.

We estimated the spectral types shown in several figures using the ($G - G_\mathrm{RP}$)--spectral type relation of E.~Mamajek, available at \url{http://www.pas.rochester.edu/~emamajek/EEM_dwarf_UBVIJHK_colors_Teff.txt}.

{A.N.~acknowledges support provided by the NSF through grant 2138089. M.A.A.~acknowledges support for this work from the TESS Guest Investigator program under NASA grant 80NSSC19K0383. J.L.C.~acknowledges support provided by the NSF through grant AST-2009840. JDNJr acknowledges the financial support by CNPq funding through the grant PQ 310847/2019-2.}

Based in part on observations obtained at the Southern Astrophysical Research (SOAR) telescope, which is a joint project of the Minist\'{e}rio da Ci\^{e}ncia, Tecnologia e Inova\c{c}\~{o}es (MCTI/LNA) do Brasil, the US National Science Foundation’s NOIRLab, the University of North Carolina at Chapel Hill (UNC), and Michigan State University (MSU).

Based in part on observations obtained at the international Gemini Observatory, a program of NSF’s NOIRLab, which is managed by the Association of Universities for Research in Astronomy (AURA) under a cooperative agreement with the National Science Foundation on behalf of the Gemini Observatory partnership: the National Science Foundation (United States), National Research Council (Canada), Agencia Nacional de Investigaci\'{o}n y Desarrollo (Chile), Ministerio de Ciencia, Tecnolog\'{i}a e Innovaci\'{o}n (Argentina), Minist\'{e}rio da Ci\^{e}ncia, Tecnologia, Inova\c{c}\~{o}es e Comunica\c{c}\~{o}es (Brazil), and Korea Astronomy and Space Science Institute (Republic of Korea).

Based on data obtained from the ESO Science Archive Facility under request number 639591.

This work has made use of data from the European Space Agency (ESA) mission Gaia,\footnote{\url{https://www.cosmos.esa.int/gaia}} processed by the Gaia Data Processing and Analysis Consortium (DPAC)\footnote{\url{https://www.cosmos.esa.int/web/gaia/dpac/consortium}}. Funding for the DPAC has been provided by national institutions, in particular the institutions participating in the Gaia Multilateral Agreement. This work also includes data collected by the TESS mission, which are publicly available from the Mikulski Archive for Space Telescopes (MAST). Funding for the TESS mission is provided by the NASA's Science Mission Directorate. 
\end{acknowledgments}

\facilities{Gaia, TESS, SOAR, Gemini-South}

\software{Astropy \citep{Astropy2013, Astropy2018};
          Goodman HTS Pipeline v. 1.3.4 \citep{GoodmanPipeline}, which uses DCR for cosmic rays identification and removal \citep{Pych2004}; 
          GMOS Data Reduction Cookbook v. 1.2.2 \citep{GMOSPipeline}.
}

\bibliography{ms.bib}{}

\begin{thebibliography}{}
\expandafter\ifx\csname natexlab\endcsname\relax\def\natexlab#1{#1}\fi
\providecommand{\url}[1]{\href{#1}{#1}}
\providecommand{\dodoi}[1]{doi:~\href{http://doi.org/#1}{\nolinkurl{#1}}}
\providecommand{\doeprint}[1]{\href{http://ascl.net/#1}{\nolinkurl{http://ascl.net/#1}}}
\providecommand{\doarXiv}[1]{\href{https://arxiv.org/abs/#1}{\nolinkurl{https://arxiv.org/abs/#1}}}

\bibitem[{{Ag{\"u}eros} {et~al.}(2018){Ag{\"u}eros}, {Bowsher}, {Bochanski},
  {Cargile}, {Covey}, {Douglas}, {Kraus}, {Kundert}, {Law}, {Ahmadi}, \&
  {Arce}}]{agueros2018}
{Ag{\"u}eros}, M.~A., {Bowsher}, E.~C., {Bochanski}, J.~J., {et~al.} 2018,
  \apj, 862, 33, \dodoi{10.3847/1538-4357/aac6ed}

\bibitem[{{Andrews} {et~al.}(2021){Andrews}, {Curtis}, {Chanam{\'e}},
  {Ag{\"u}eros}, {Schuler}, {Kounkel}, \& {Covey}}]{Andrews2021}
{Andrews}, J.~J., {Curtis}, J.~L., {Chanam{\'e}}, J., {et~al.} 2021, arXiv
  e-prints, arXiv:2110.06278.
\newblock \doarXiv{2110.06278}

\bibitem[{{Astropy Collaboration} {et~al.}(2013){Astropy Collaboration},
  {Robitaille}, {Tollerud}, {Greenfield}, {Droettboom}, {Bray}, {Aldcroft},
  {Davis}, {Ginsburg}, {Price-Whelan}, {Kerzendorf}, {Conley}, {Crighton},
  {Barbary}, {Muna}, {Ferguson}, {Grollier}, {Parikh}, {Nair}, {Unther},
  {Deil}, {Woillez}, {Conseil}, {Kramer}, {Turner}, {Singer}, {Fox}, {Weaver},
  {Zabalza}, {Edwards}, {Azalee Bostroem}, {Burke}, {Casey}, {Crawford},
  {Dencheva}, {Ely}, {Jenness}, {Labrie}, {Lim}, {Pierfederici}, {Pontzen},
  {Ptak}, {Refsdal}, {Servillat}, \& {Streicher}}]{Astropy2013}
{Astropy Collaboration}, {Robitaille}, T.~P., {Tollerud}, E.~J., {et~al.} 2013,
  \aap, 558, A33, \dodoi{10.1051/0004-6361/201322068}

\bibitem[{{Astropy Collaboration} {et~al.}(2018){Astropy Collaboration},
  {Price-Whelan}, {Sip{\H{o}}cz}, {G{\"u}nther}, {Lim}, {Crawford}, {Conseil},
  {Shupe}, {Craig}, {Dencheva}, {Ginsburg}, {VanderPlas}, {Bradley},
  {P{\'e}rez-Su{\'a}rez}, {de Val-Borro}, {Aldcroft}, {Cruz}, {Robitaille},
  {Tollerud}, {Ardelean}, {Babej}, {Bach}, {Bachetti}, {Bakanov}, {Bamford},
  {Barentsen}, {Barmby}, {Baumbach}, {Berry}, {Biscani}, {Boquien}, {Bostroem},
  {Bouma}, {Brammer}, {Bray}, {Breytenbach}, {Buddelmeijer}, {Burke},
  {Calderone}, {Cano Rodr{\'\i}guez}, {Cara}, {Cardoso}, {Cheedella}, {Copin},
  {Corrales}, {Crichton}, {D'Avella}, {Deil}, {Depagne}, {Dietrich}, {Donath},
  {Droettboom}, {Earl}, {Erben}, {Fabbro}, {Ferreira}, {Finethy}, {Fox},
  {Garrison}, {Gibbons}, {Goldstein}, {Gommers}, {Greco}, {Greenfield},
  {Groener}, {Grollier}, {Hagen}, {Hirst}, {Homeier}, {Horton}, {Hosseinzadeh},
  {Hu}, {Hunkeler}, {Ivezi{\'c}}, {Jain}, {Jenness}, {Kanarek}, {Kendrew},
  {Kern}, {Kerzendorf}, {Khvalko}, {King}, {Kirkby}, {Kulkarni}, {Kumar},
  {Lee}, {Lenz}, {Littlefair}, {Ma}, {Macleod}, {Mastropietro}, {McCully},
  {Montagnac}, {Morris}, {Mueller}, {Mumford}, {Muna}, {Murphy}, {Nelson},
  {Nguyen}, {Ninan}, {N{\"o}the}, {Ogaz}, {Oh}, {Parejko}, {Parley}, {Pascual},
  {Patil}, {Patil}, {Plunkett}, {Prochaska}, {Rastogi}, {Reddy Janga},
  {Sabater}, {Sakurikar}, {Seifert}, {Sherbert}, {Sherwood-Taylor}, {Shih},
  {Sick}, {Silbiger}, {Singanamalla}, {Singer}, {Sladen}, {Sooley},
  {Sornarajah}, {Streicher}, {Teuben}, {Thomas}, {Tremblay}, {Turner},
  {Terr{\'o}n}, {van Kerkwijk}, {de la Vega}, {Watkins}, {Weaver}, {Whitmore},
  {Woillez}, {Zabalza}, \& {Astropy Contributors}}]{Astropy2018}
{Astropy Collaboration}, {Price-Whelan}, A.~M., {Sip{\H{o}}cz}, B.~M., {et~al.}
  2018, \aj, 156, 123, \dodoi{10.3847/1538-3881/aabc4f}

\bibitem[{{Astudillo-Defru} {et~al.}(2017){Astudillo-Defru}, {Delfosse},
  {Bonfils}, {Forveille}, {Lovis}, \& {Rameau}}]{Astudillo2017}
{Astudillo-Defru}, N., {Delfosse}, X., {Bonfils}, X., {et~al.} 2017, \aap, 600,
  A13, \dodoi{10.1051/0004-6361/201527078}

\bibitem[{{Bailer-Jones} {et~al.}(2021){Bailer-Jones}, {Rybizki}, {Fouesneau},
  {Demleitner}, \& {Andrae}}]{Bailer-Jones2021}
{Bailer-Jones}, C.~A.~L., {Rybizki}, J., {Fouesneau}, M., {Demleitner}, M., \&
  {Andrae}, R. 2021, {VizieR Online Data Catalog: Distances to 1.47 billion
  stars in Gaia EDR3 (Bailer-Jones+, 2021)}.
\newblock \url{https://cdsarc.cds.unistra.fr/viz-bin/cat/I/352}

\bibitem[{{Basri} {et~al.}(2005){Basri}, {Borucki}, \& {Koch}}]{kepler2005}
{Basri}, G., {Borucki}, W.~J., \& {Koch}, D. 2005, \nar, 49, 478,
  \dodoi{10.1016/j.newar.2005.08.026}

\bibitem[{{Bellm} {et~al.}(2019){Bellm}, {Kulkarni}, {Graham}, {Dekany},
  {Smith}, {Riddle}, {Masci}, {Helou}, {Prince}, {Adams}, {Barbarino},
  {Barlow}, {Bauer}, {Beck}, {Belicki}, {Biswas}, {Blagorodnova}, {Bodewits},
  {Bolin}, {Brinnel}, {Brooke}, {Bue}, {Bulla}, {Burruss}, {Cenko}, {Chang},
  {Connolly}, {Coughlin}, {Cromer}, {Cunningham}, {De}, {Delacroix}, {Desai},
  {Duev}, {Eadie}, {Farnham}, {Feeney}, {Feindt}, {Flynn}, {Franckowiak},
  {Frederick}, {Fremling}, {Gal-Yam}, {Gezari}, {Giomi}, {Goldstein},
  {Golkhou}, {Goobar}, {Groom}, {Hacopians}, {Hale}, {Henning}, {Ho}, {Hover},
  {Howell}, {Hung}, {Huppenkothen}, {Imel}, {Ip}, {Ivezi{\'c}}, {Jackson},
  {Jones}, {Juric}, {Kasliwal}, {Kaspi}, {Kaye}, {Kelley}, {Kowalski},
  {Kramer}, {Kupfer}, {Landry}, {Laher}, {Lee}, {Lin}, {Lin}, {Lunnan},
  {Giomi}, {Mahabal}, {Mao}, {Miller}, {Monkewitz}, {Murphy}, {Ngeow},
  {Nordin}, {Nugent}, {Ofek}, {Patterson}, {Penprase}, {Porter}, {Rauch},
  {Rebbapragada}, {Reiley}, {Rigault}, {Rodriguez}, {van Roestel}, {Rusholme},
  {van Santen}, {Schulze}, {Shupe}, {Singer}, {Soumagnac}, {Stein}, {Surace},
  {Sollerman}, {Szkody}, {Taddia}, {Terek}, {Van Sistine}, {van Velzen},
  {Vestrand}, {Walters}, {Ward}, {Ye}, {Yu}, {Yan}, \& {Zolkower}}]{Bellm2019}
{Bellm}, E.~C., {Kulkarni}, S.~R., {Graham}, M.~J., {et~al.} 2019, \pasp, 131,
  018002, \dodoi{10.1088/1538-3873/aaecbe}

\bibitem[{{Belokurov} {et~al.}(2020){Belokurov}, {Penoyre}, {Oh}, {Iorio},
  {Hodgkin}, {Evans}, {Everall}, {Koposov}, {Tout}, {Izzard}, {Clarke}, \&
  {Brown}}]{Belokurov2020}
{Belokurov}, V., {Penoyre}, Z., {Oh}, S., {et~al.} 2020, \mnras, 496, 1922,
  \dodoi{10.1093/mnras/staa1522}

\bibitem[{{Berta} {et~al.}(2012){Berta}, {Irwin}, {Charbonneau}, {Burke}, \&
  {Falco}}]{Berta2012}
{Berta}, Z.~K., {Irwin}, J., {Charbonneau}, D., {Burke}, C.~J., \& {Falco},
  E.~E. 2012, \aj, 144, 145, \dodoi{10.1088/0004-6256/144/5/145}

\bibitem[{{Biazzo} {et~al.}(2012){Biazzo}, {Alcal{\'a}}, {Covino}, {Frasca},
  {Getman}, \& {Spezzi}}]{Biazzo2012}
{Biazzo}, K., {Alcal{\'a}}, J.~M., {Covino}, E., {et~al.} 2012, \aap, 547,
  A104, \dodoi{10.1051/0004-6361/201219680}

\bibitem[{{Borucki} {et~al.}(2010){Borucki}, {Koch}, {Basri}, {Batalha},
  {Brown}, {Caldwell}, {Caldwell}, {Christensen-Dalsgaard}, {Cochran},
  {DeVore}, {Dunham}, {Dupree}, {Gautier}, {Geary}, {Gilliland}, {Gould},
  {Howell}, {Jenkins}, {Kondo}, {Latham}, {Marcy}, {Meibom}, {Kjeldsen},
  {Lissauer}, {Monet}, {Morrison}, {Sasselov}, {Tarter}, {Boss}, {Brownlee},
  {Owen}, {Buzasi}, {Charbonneau}, {Doyle}, {Fortney}, {Ford}, {Holman},
  {Seager}, {Steffen}, {Welsh}, {Rowe}, {Anderson}, {Buchhave}, {Ciardi},
  {Walkowicz}, {Sherry}, {Horch}, {Isaacson}, {Everett}, {Fischer}, {Torres},
  {Johnson}, {Endl}, {MacQueen}, {Bryson}, {Dotson}, {Haas}, {Kolodziejczak},
  {Van Cleve}, {Chandrasekaran}, {Twicken}, {Quintana}, {Clarke}, {Allen},
  {Li}, {Wu}, {Tenenbaum}, {Verner}, {Bruhweiler}, {Barnes}, \&
  {Prsa}}]{borucki2010}
{Borucki}, W.~J., {Koch}, D., {Basri}, G., {et~al.} 2010, Science, 327, 977,
  \dodoi{10.1126/science.1185402}

\bibitem[{{Bradley} {et~al.}(2016){Bradley}, {Sipocz}, {Robitaille},
  {Tollerud}, {Vin{\'\i}cius}, {Deil}, {Barbary}, {G{\"u}nther}, {Cara},
  {Droettboom}, {Bostroem}, {Bray}, {Andersen Bratholm}, {Pickering}, {Craig},
  {Barentsen}, {Pascual}, {adonath}, {Greco}, {Kerzendorf}, {StuartLittlefair},
  {Ferreira}, {D'Eugenio}, \& {Weaver}}]{Bradley2016}
{Bradley}, L., {Sipocz}, B., {Robitaille}, T., {et~al.} 2016,
  {Astropy/Photutils: V0.3}, v0.3,  Zenodo, \dodoi{10.5281/zenodo.164986}

\bibitem[{{Brasseur} {et~al.}(2019){Brasseur}, {Phillip}, {Fleming},
  {Mullally}, \& {White}}]{Brasseur2019}
{Brasseur}, C.~E., {Phillip}, C., {Fleming}, S.~W., {Mullally}, S.~E., \&
  {White}, R.~L. 2019, {Astrocut: Tools for creating cutouts of TESS images}.
\newblock \doeprint{1905.007}

\bibitem[{{Chabrier} \& {Baraffe}(1997)}]{Chabrier1997}
{Chabrier}, G., \& {Baraffe}, I. 1997, \aap, 327, 1039.
\newblock \doarXiv{astro-ph/9704118}

\bibitem[{{Claytor} {et~al.}(2021){Claytor}, {van Saders}, {Llama}, {Sadowski},
  {Quach}, \& {Avallone}}]{Claytor2021}
{Claytor}, Z.~R., {van Saders}, J.~L., {Llama}, J., {et~al.} 2021, arXiv
  e-prints, arXiv:2104.14566.
\newblock \doarXiv{2104.14566}

\bibitem[{{Clemens} {et~al.}(2004){Clemens}, {Crain}, \&
  {Anderson}}]{Clemens2004}
{Clemens}, J.~C., {Crain}, J.~A., \& {Anderson}, R. 2004, in Society of
  Photo-Optical Instrumentation Engineers (SPIE) Conference Series, Vol. 5492,
  Ground-based Instrumentation for Astronomy, ed. A.~F.~M. {Moorwood} \&
  M.~{Iye}, 331--340, \dodoi{10.1117/12.550069}

\bibitem[{{Collier Cameron} \& {Jianke}(1994)}]{Cameron1994}
{Collier Cameron}, A., \& {Jianke}, L. 1994, \mnras, 269, 1099,
  \dodoi{10.1093/mnras/269.4.1099}

\bibitem[{{Delfosse} {et~al.}(1998){Delfosse}, {Forveille}, {Perrier}, \&
  {Mayor}}]{Delfosse1998}
{Delfosse}, X., {Forveille}, T., {Perrier}, C., \& {Mayor}, M. 1998, \aap, 331,
  581

\bibitem[{{Douglas} {et~al.}(2017){Douglas}, {Ag{\"u}eros}, {Covey}, \&
  {Kraus}}]{Douglas2017}
{Douglas}, S.~T., {Ag{\"u}eros}, M.~A., {Covey}, K.~R., \& {Kraus}, A. 2017,
  \apj, 842, 83, \dodoi{10.3847/1538-4357/aa6e52}

\bibitem[{{Douglas} {et~al.}(2019){Douglas}, {Curtis}, {Ag{\"u}eros},
  {Cargile}, {Brewer}, {Meibom}, \& {Jansen}}]{Douglas2019}
{Douglas}, S.~T., {Curtis}, J.~L., {Ag{\"u}eros}, M.~A., {et~al.} 2019, \apj,
  879, 100, \dodoi{10.3847/1538-4357/ab2468}

\bibitem[{{Douglas} {et~al.}(2014){Douglas}, {Ag{\"u}eros}, {Covey}, {Bowsher},
  {Bochanski}, {Cargile}, {Kraus}, {Law}, {Lemonias}, {Arce}, {Fierroz}, \&
  {Kundert}}]{Douglas2014}
{Douglas}, S.~T., {Ag{\"u}eros}, M.~A., {Covey}, K.~R., {et~al.} 2014, \apj,
  795, 161, \dodoi{10.1088/0004-637X/795/2/161}

\bibitem[{{Dressing} \& {Charbonneau}(2015)}]{Dressing2015}
{Dressing}, C.~D., \& {Charbonneau}, D. 2015, \apj, 807, 45,
  \dodoi{10.1088/0004-637X/807/1/45}

\bibitem[{{Dymock} \& {Miles}(2009)}]{Dymock2009}
{Dymock}, R., \& {Miles}, R. 2009, Journal of the British Astronomical
  Association, 119, 149.
\newblock \doarXiv{1006.4017}

\bibitem[{{France} {et~al.}(2020){France}, {Duvvuri}, {Egan}, {Koskinen},
  {Wilson}, {Youngblood}, {Froning}, {Brown}, {Alvarado-G{\'o}mez},
  {Berta-Thompson}, {Drake}, {Garraffo}, {Kaltenegger}, {Kowalski}, {Linsky},
  {Loyd}, {Mauas}, {Miguel}, {Pineda}, {Rugheimer}, {Schneider}, {Tian}, \&
  {Vieytes}}]{France2020}
{France}, K., {Duvvuri}, G., {Egan}, H., {et~al.} 2020, \aj, 160, 237,
  \dodoi{10.3847/1538-3881/abb465}

\bibitem[{{Gaia Collaboration} {et~al.}(2021){Gaia Collaboration}, {Brown},
  {Vallenari}, {Prusti}, {de Bruijne}, {Babusiaux}, {Biermann}, {Creevey},
  {Evans}, {Eyer}, {Hutton}, {Jansen}, {Jordi}, {Klioner}, {Lammers},
  {Lindegren}, {Luri}, {Mignard}, {Panem}, {Pourbaix}, {Randich}, {Sartoretti},
  {Soubiran}, {Walton}, {Arenou}, {Bailer-Jones}, {Bastian}, {Cropper},
  {Drimmel}, {Katz}, {Lattanzi}, {van Leeuwen}, {Bakker}, {Cacciari},
  {Casta{\~n}eda}, {De Angeli}, {Ducourant}, {Fabricius}, {Fouesneau},
  {Fr{\'e}mat}, {Guerra}, {Guerrier}, {Guiraud}, {Jean-Antoine Piccolo},
  {Masana}, {Messineo}, {Mowlavi}, {Nicolas}, {Nienartowicz}, {Pailler},
  {Panuzzo}, {Riclet}, {Roux}, {Seabroke}, {Sordo}, {Tanga}, {Th{\'e}venin},
  {Gracia-Abril}, {Portell}, {Teyssier}, {Altmann}, {Andrae}, {Bellas-Velidis},
  {Benson}, {Berthier}, {Blomme}, {Brugaletta}, {Burgess}, {Busso}, {Carry},
  {Cellino}, {Cheek}, {Clementini}, {Damerdji}, {Davidson}, {Delchambre},
  {Dell'Oro}, {Fern{\'a}ndez-Hern{\'a}ndez}, {Galluccio}, {Garc{\'\i}a-Lario},
  {Garcia-Reinaldos}, {Gonz{\'a}lez-N{\'u}{\~n}ez}, {Gosset}, {Haigron},
  {Halbwachs}, {Hambly}, {Harrison}, {Hatzidimitriou}, {Heiter},
  {Hern{\'a}ndez}, {Hestroffer}, {Hodgkin}, {Holl}, {Jan{\ss}en}, {Jevardat de
  Fombelle}, {Jordan}, {Krone-Martins}, {Lanzafame}, {L{\"o}ffler}, {Lorca},
  {Manteiga}, {Marchal}, {Marrese}, {Moitinho}, {Mora}, {Muinonen}, {Osborne},
  {Pancino}, {Pauwels}, {Petit}, {Recio-Blanco}, {Richards}, {Riello},
  {Rimoldini}, {Robin}, {Roegiers}, {Rybizki}, {Sarro}, {Siopis}, {Smith},
  {Sozzetti}, {Ulla}, {Utrilla}, {van Leeuwen}, {van Reeven}, {Abbas}, {Abreu
  Aramburu}, {Accart}, {Aerts}, {Aguado}, {Ajaj}, {Altavilla}, {{\'A}lvarez},
  {{\'A}lvarez Cid-Fuentes}, {Alves}, {Anderson}, {Anglada Varela}, {Antoja},
  {Audard}, {Baines}, {Baker}, {Balaguer-N{\'u}{\~n}ez}, {Balbinot}, {Balog},
  {Barache}, {Barbato}, {Barros}, {Barstow}, {Bartolom{\'e}}, {Bassilana},
  {Bauchet}, {Baudesson-Stella}, {Becciani}, {Bellazzini}, {Bernet}, {Bertone},
  {Bianchi}, {Blanco-Cuaresma}, {Boch}, {Bombrun}, {Bossini}, {Bouquillon},
  {Bragaglia}, {Bramante}, {Breedt}, {Bressan}, {Brouillet}, {Bucciarelli},
  {Burlacu}, {Busonero}, {Butkevich}, {Buzzi}, {Caffau}, {Cancelliere},
  {C{\'a}novas}, {Cantat-Gaudin}, {Carballo}, {Carlucci}, {Carnerero},
  {Carrasco}, {Casamiquela}, {Castellani}, {Castro-Ginard}, {Castro Sampol},
  {Chaoul}, {Charlot}, {Chemin}, {Chiavassa}, {Cioni}, {Comoretto}, {Cooper},
  {Cornez}, {Cowell}, {Crifo}, {Crosta}, {Crowley}, {Dafonte}, {Dapergolas},
  {David}, {David}, {de Laverny}, {De Luise}, {De March}, {De Ridder}, {de
  Souza}, {de Teodoro}, {de Torres}, {del Peloso}, {del Pozo}, {Delbo},
  {Delgado}, {Delgado}, {Delisle}, {Di Matteo}, {Diakite}, {Diener},
  {Distefano}, {Dolding}, {Eappachen}, {Edvardsson}, {Enke}, {Esquej}, {Fabre},
  {Fabrizio}, {Faigler}, {Fedorets}, {Fernique}, {Fienga}, {Figueras},
  {Fouron}, {Fragkoudi}, {Fraile}, {Franke}, {Gai}, {Garabato},
  {Garcia-Gutierrez}, {Garc{\'\i}a-Torres}, {Garofalo}, {Gavras}, {Gerlach},
  {Geyer}, {Giacobbe}, {Gilmore}, {Girona}, {Giuffrida}, {Gomel}, {Gomez},
  {Gonzalez-Santamaria}, {Gonz{\'a}lez-Vidal}, {Granvik},
  {Guti{\'e}rrez-S{\'a}nchez}, {Guy}, {Hauser}, {Haywood}, {Helmi}, {Hidalgo},
  {Hilger}, {H{\l}adczuk}, {Hobbs}, {Holland}, {Huckle}, {Jasniewicz},
  {Jonker}, {Juaristi Campillo}, {Julbe}, {Karbevska}, {Kervella}, {Khanna},
  {Kochoska}, {Kontizas}, {Kordopatis}, {Korn}, {Kostrzewa-Rutkowska},
  {Kruszy{\'n}ska}, {Lambert}, {Lanza}, {Lasne}, {Le Campion}, {Le Fustec},
  {Lebreton}, {Lebzelter}, {Leccia}, {Leclerc}, {Lecoeur-Taibi}, {Liao},
  {Licata}, {Lindstr{\o}m}, {Lister}, {Livanou}, {Lobel}, {Madrero Pardo},
  {Managau}, {Mann}, {Marchant}, {Marconi}, {Marcos Santos}, {Marinoni},
  {Marocco}, {Marshall}, {Martin Polo}, {Mart{\'\i}n-Fleitas}, {Masip},
  {Massari}, {Mastrobuono-Battisti}, {Mazeh}, {McMillan}, {Messina},
  {Michalik}, {Millar}, {Mints}, {Molina}, {Molinaro}, {Moln{\'a}r},
  {Montegriffo}, {Mor}, {Morbidelli}, {Morel}, {Morris}, {Mulone}, {Munoz},
  {Muraveva}, {Murphy}, {Musella}, {Noval}, {Ord{\'e}novic}, {Orr{\`u}},
  {Osinde}, {Pagani}, {Pagano}, {Palaversa}, {Palicio}, {Panahi}, {Pawlak},
  {Pe{\~n}alosa Esteller}, {Penttil{\"a}}, {Piersimoni}, {Pineau}, {Plachy},
  {Plum}, {Poggio}, {Poretti}, {Poujoulet}, {Pr{\v{s}}a}, {Pulone}, {Racero},
  {Ragaini}, {Rainer}, {Raiteri}, {Rambaux}, {Ramos}, {Ramos-Lerate}, {Re
  Fiorentin}, {Regibo}, {Reyl{\'e}}, {Ripepi}, {Riva}, {Rixon}, {Robichon},
  {Robin}, {Roelens}, {Rohrbasser}, {Romero-G{\'o}mez}, {Rowell}, {Royer},
  {Rybicki}, {Sadowski}, {Sagrist{\`a} Sell{\'e}s}, {Sahlmann}, {Salgado},
  {Salguero}, {Samaras}, {Sanchez Gimenez}, {Sanna}, {Santove{\~n}a},
  {Sarasso}, {Schultheis}, {Sciacca}, {Segol}, {Segovia}, {S{\'e}gransan},
  {Semeux}, {Shahaf}, {Siddiqui}, {Siebert}, {Siltala}, {Slezak}, {Smart},
  {Solano}, {Solitro}, {Souami}, {Souchay}, {Spagna}, {Spoto}, {Steele},
  {Steidelm{\"u}ller}, {Stephenson}, {S{\"u}veges}, {Szabados}, {Szegedi-Elek},
  {Taris}, {Tauran}, {Taylor}, {Teixeira}, {Thuillot}, {Tonello}, {Torra},
  {Torra}, {Turon}, {Unger}, {Vaillant}, {van Dillen}, {Vanel}, {Vecchiato},
  {Viala}, {Vicente}, {Voutsinas}, {Weiler}, {Wevers}, {Wyrzykowski}, {Yoldas},
  {Yvard}, {Zhao}, {Zorec}, {Zucker}, {Zurbach}, \& {Zwitter}}]{Gaia2020}
{Gaia Collaboration}, {Brown}, A.~G.~A., {Vallenari}, A., {et~al.} 2021, \aap,
  649, A1, \dodoi{10.1051/0004-6361/202039657}

\bibitem[{{Garraffo} {et~al.}(2015){Garraffo}, {Drake}, \&
  {Cohen}}]{Garraffo2015}
{Garraffo}, C., {Drake}, J.~J., \& {Cohen}, O. 2015, \apj, 813, 40,
  \dodoi{10.1088/0004-637X/813/1/40}

\bibitem[{{Garraffo} {et~al.}(2017){Garraffo}, {Drake}, {Cohen},
  {Alvarado-G{\'o}mez}, \& {Moschou}}]{garraffo2017}
{Garraffo}, C., {Drake}, J.~J., {Cohen}, O., {Alvarado-G{\'o}mez}, J.~D., \&
  {Moschou}, S.~P. 2017, \apjl, 843, L33, \dodoi{10.3847/2041-8213/aa79ed}

\bibitem[{{Gillon} {et~al.}(2017){Gillon}, {Triaud}, {Demory}, {Jehin}, {Agol},
  {Deck}, {Lederer}, {de Wit}, {Burdanov}, {Ingalls}, {Bolmont}, {Leconte},
  {Raymond}, {Selsis}, {Turbet}, {Barkaoui}, {Burgasser}, {Burleigh}, {Carey},
  {Chaushev}, {Copperwheat}, {Delrez}, {Fernandes}, {Holdsworth}, {Kotze}, {Van
  Grootel}, {Almleaky}, {Benkhaldoun}, {Magain}, \& {Queloz}}]{Gillon2017}
{Gillon}, M., {Triaud}, A.~H.~M.~J., {Demory}, B.-O., {et~al.} 2017, \nat, 542,
  456, \dodoi{10.1038/nature21360}

\bibitem[{Gimeno {et~al.}(2016)Gimeno, Roth, Chiboucas, Hibon, Boucher, White,
  Rippa, Labrie, Turner, Hanna, Lazo, Pérez, Rogers, Rojas, Placco, \&
  Murowinski}]{gmoss}
Gimeno, G., Roth, K., Chiboucas, K., {et~al.} 2016, in Ground-based and
  Airborne Instrumentation for Astronomy VI, ed. C.~J. Evans, L.~Simard, \&
  H.~Takami, Vol. 9908, International Society for Optics and Photonics (SPIE),
  872 -- 885, \dodoi{10.1117/12.2233883}

\bibitem[{{Ginsburg} \& {Mirocha}(2011)}]{Ginsburg2011}
{Ginsburg}, A., \& {Mirocha}, J. 2011, {PySpecKit: Python Spectroscopic
  Toolkit}, Astrophysics Source Code Library.
\newblock \doeprint{1109.001}

\bibitem[{{Glazier} {et~al.}(2020){Glazier}, {Howard}, {Corbett}, {Law},
  {Ratzloff}, {Fors}, \& {del Ser}}]{Glazier2020}
{Glazier}, A.~L., {Howard}, W.~S., {Corbett}, H., {et~al.} 2020, \apj, 900, 27,
  \dodoi{10.3847/1538-4357/aba4a6}

\bibitem[{{Gould} {et~al.}(2003){Gould}, {Pepper}, \& {DePoy}}]{Gould2003}
{Gould}, A., {Pepper}, J., \& {DePoy}, D.~L. 2003, \apj, 594, 533,
  \dodoi{10.1086/376852}

\bibitem[{{Gunning} {et~al.}(2014){Gunning}, {Schmidt}, {Davenport}, {Dhital},
  {Hawley}, \& {West}}]{Gunning2014}
{Gunning}, H.~C., {Schmidt}, S.~J., {Davenport}, J.~R.~A., {et~al.} 2014,
  \pasp, 126, 1081, \dodoi{10.1086/679329}

\bibitem[{{Haisch} {et~al.}(2001){Haisch}, {Lada}, \& {Lada}}]{Haisch2001}
{Haisch}, Karl~E., J., {Lada}, E.~A., \& {Lada}, C.~J. 2001, \apjl, 553, L153,
  \dodoi{10.1086/320685}

\bibitem[{{Hattori} {et~al.}(2021){Hattori}, {Foreman-Mackey}, {Hogg},
  {Montet}, {Angus}, {Pritchard}, {Curtis}, \& {Sch{\"o}lkopf}}]{Hattori2021}
{Hattori}, S., {Foreman-Mackey}, D., {Hogg}, D.~W., {et~al.} 2021, arXiv
  e-prints, arXiv:2106.15063.
\newblock \doarXiv{2106.15063}

\bibitem[{{Henden} {et~al.}(2016){Henden}, {Templeton}, {Terrell}, {Smith},
  {Levine}, \& {Welch}}]{Henden2016}
{Henden}, A.~A., {Templeton}, M., {Terrell}, D., {et~al.} 2016, {VizieR Online
  Data Catalog: AAVSO Photometric All Sky Survey (APASS) DR9 (Henden+, 2016)}.
\newblock \url{https://cdsarc.cds.unistra.fr/viz-bin/cat/II/336}

\bibitem[{{Howard} {et~al.}(2020){Howard}, {Corbett}, {Law}, {Ratzloff},
  {Galliher}, {Glazier}, {Fors}, {del Ser}, \& {Haislip}}]{Howard2020}
{Howard}, W.~S., {Corbett}, H., {Law}, N.~M., {et~al.} 2020, \apj, 895, 140,
  \dodoi{10.3847/1538-4357/ab9081}

\bibitem[{{Howard} {et~al.}(2021){Howard}, {Teske}, {Corbett}, {Law}, {Wang},
  {Ratzloff}, {Galliher}, {Gonzalez}, {Soto}, {Glazier}, \&
  {Haislip}}]{Howard2021}
{Howard}, W.~S., {Teske}, J., {Corbett}, H., {et~al.} 2021, \aj, 162, 147,
  \dodoi{10.3847/1538-3881/ac0fe3}

\bibitem[{{Husser} {et~al.}(2013){Husser}, {Wende-von Berg}, {Dreizler},
  {Homeier}, {Reiners}, {Barman}, \& {Hauschildt}}]{Husser2013}
{Husser}, T.~O., {Wende-von Berg}, S., {Dreizler}, S., {et~al.} 2013, \aap,
  553, A6, \dodoi{10.1051/0004-6361/201219058}

\bibitem[{{Jackson} \& {Jeffries}(2010)}]{Jackson2010}
{Jackson}, R.~J., \& {Jeffries}, R.~D. 2010, \mnras, 407, 465,
  \dodoi{10.1111/j.1365-2966.2010.16917.x}

\bibitem[{{Jorissen}(2019)}]{Jorissen2019}
{Jorissen}, A. 2019, \memsai, 90, 395

\bibitem[{{Kiman} {et~al.}(2021){Kiman}, {Faherty}, {Cruz}, {Gagn{\'e}},
  {Angus}, {Schmidt}, {Mann}, {Bardalez Gagliuffi}, \& {Rice}}]{Kiman2021}
{Kiman}, R., {Faherty}, J.~K., {Cruz}, K.~L., {et~al.} 2021, \aj, 161, 277,
  \dodoi{10.3847/1538-3881/abf561}

\bibitem[{{Kraus} {et~al.}(2016){Kraus}, {Ireland}, {Huber}, {Mann}, \&
  {Dupuy}}]{Kraus2016}
{Kraus}, A.~L., {Ireland}, M.~J., {Huber}, D., {Mann}, A.~W., \& {Dupuy}, T.~J.
  2016, \aj, 152, 8, \dodoi{10.3847/0004-6256/152/1/8}

\bibitem[{{Lee} {et~al.}(2020){Lee}, {Song}, \& {Murphy}}]{Lee2020}
{Lee}, J., {Song}, I., \& {Murphy}, S. 2020, \mnras, 494, 62,
  \dodoi{10.1093/mnras/staa689}

\bibitem[{{Mann} {et~al.}(2019){Mann}, {Dupuy}, {Kraus}, {Gaidos}, {Ansdell},
  {Ireland}, {Rizzuto}, {Hung}, {Dittmann}, {Factor}, {Feiden}, {Martinez},
  {Ru{\'\i}z-Rodr{\'\i}guez}, \& {Thao}}]{Mann2019}
{Mann}, A.~W., {Dupuy}, T., {Kraus}, A.~L., {et~al.} 2019, \apj, 871, 63,
  \dodoi{10.3847/1538-4357/aaf3bc}

\bibitem[{{Mayor} {et~al.}(2003){Mayor}, {Pepe}, {Queloz}, {Bouchy},
  {Rupprecht}, {Lo Curto}, {Avila}, {Benz}, {Bertaux}, {Bonfils}, {Dall},
  {Dekker}, {Delabre}, {Eckert}, {Fleury}, {Gilliotte}, {Gojak}, {Guzman},
  {Kohler}, {Lizon}, {Longinotti}, {Lovis}, {Megevand}, {Pasquini}, {Reyes},
  {Sivan}, {Sosnowska}, {Soto}, {Udry}, {van Kesteren}, {Weber}, \&
  {Weilenmann}}]{Mayor2003}
{Mayor}, M., {Pepe}, F., {Queloz}, D., {et~al.} 2003, The Messenger, 114, 20

\bibitem[{{Meibom} {et~al.}(2007){Meibom}, {Mathieu}, \&
  {Stassun}}]{Meibom2007}
{Meibom}, S., {Mathieu}, R.~D., \& {Stassun}, K.~G. 2007, \apjl, 665, L155,
  \dodoi{10.1086/521437}

\bibitem[{{Messina} {et~al.}(2017){Messina}, {Lanzafame}, {Malo}, {Desidera},
  {Buccino}, {Zhang}, {Artemenko}, {Millward}, \& {Hambsch}}]{Messina2017}
{Messina}, S., {Lanzafame}, A.~C., {Malo}, L., {et~al.} 2017, \aap, 607, A3,
  \dodoi{10.1051/0004-6361/201730444}

\bibitem[{{Mohanty} \& {Basri}(2003)}]{Mohanty2003}
{Mohanty}, S., \& {Basri}, G. 2003, \apj, 583, 451, \dodoi{10.1086/345097}

\bibitem[{{Morgan} {et~al.}(2012){Morgan}, {West}, {Garc{\'e}s}, {Catal{\'a}n},
  {Dhital}, {Fuchs}, \& {Silvestri}}]{Morgan2012}
{Morgan}, D.~P., {West}, A.~A., {Garc{\'e}s}, A., {et~al.} 2012, \aj, 144, 93,
  \dodoi{10.1088/0004-6256/144/4/93}

\bibitem[{{Muirhead} {et~al.}(2015){Muirhead}, {Mann}, {Vanderburg}, {Morton},
  {Kraus}, {Ireland}, {Swift}, {Feiden}, {Gaidos}, \& {Gazak}}]{Muirhead2015}
{Muirhead}, P.~S., {Mann}, A.~W., {Vanderburg}, A., {et~al.} 2015, \apj, 801,
  18, \dodoi{10.1088/0004-637X/801/1/18}

\bibitem[{{Muzerolle} {et~al.}(1998){Muzerolle}, {Hartmann}, \&
  {Calvet}}]{Muzerolle1998}
{Muzerolle}, J., {Hartmann}, L., \& {Calvet}, N. 1998, \aj, 116, 2965,
  \dodoi{10.1086/300636}

\bibitem[{{Newton} {et~al.}(2017){Newton}, {Irwin}, {Charbonneau}, {Berlind},
  {Calkins}, \& {Mink}}]{Newton2017}
{Newton}, E.~R., {Irwin}, J., {Charbonneau}, D., {et~al.} 2017, \apj, 834, 85,
  \dodoi{10.3847/1538-4357/834/1/85}

\bibitem[{{Noyes} {et~al.}(1984){Noyes}, {Weiss}, \& {Vaughan}}]{Noyes1984}
{Noyes}, R.~W., {Weiss}, N.~O., \& {Vaughan}, A.~H. 1984, \apj, 287, 769,
  \dodoi{10.1086/162735}

\bibitem[{{N{\'u}{\~n}ez} {et~al.}(2022){N{\'u}{\~n}ez}, {Douglas}, {Alam}, \&
  {DeLaurentiis}}]{Alam2016}
{N{\'u}{\~n}ez}, A., {Douglas}, S., {Alam}, M., \& {DeLaurentiis}, S. 2022,
  PHEW: PytHon Equivalent Widths, v2.0,  Zenodo, \dodoi{10.5281/zenodo.6422571}

\bibitem[{{N{\'u}{\~n}ez} {et~al.}(2015){N{\'u}{\~n}ez}, {Ag{\"u}eros},
  {Covey}, {Hartman}, {Kraus}, {Bowsher}, {Douglas}, {L{\'o}pez-Morales},
  {Pooley}, {Posselt}, {Saar}, \& {West}}]{Nunez2015}
{N{\'u}{\~n}ez}, A., {Ag{\"u}eros}, M.~A., {Covey}, K.~R., {et~al.} 2015, \apj,
  809, 161, \dodoi{10.1088/0004-637X/809/2/161}

\bibitem[{{Pych}(2004)}]{Pych2004}
{Pych}, W. 2004, \pasp, 116, 148, \dodoi{10.1086/381786}

\bibitem[{{Rampalli} {et~al.}(2021){Rampalli}, {Ag{\"u}eros}, {Curtis},
  {Douglas}, {N{\'u}{\~n}ez}, {Cargile}, {Covey}, {Gosnell}, {Kraus}, {Law}, \&
  {Mann}}]{Rampalli2021}
{Rampalli}, R., {Ag{\"u}eros}, M.~A., {Curtis}, J.~L., {et~al.} 2021, \apj,
  921, 167, \dodoi{10.3847/1538-4357/ac0c1e}

\bibitem[{{Ratzloff} {et~al.}(2019){Ratzloff}, {Law}, {Fors}, {Corbett},
  {Howard}, {del Ser}, \& {Haislip}}]{Ratzloff2019}
{Ratzloff}, J.~K., {Law}, N.~M., {Fors}, O., {et~al.} 2019, \pasp, 131, 075001,
  \dodoi{10.1088/1538-3873/ab19d0}

\bibitem[{{Rebull} {et~al.}(2006){Rebull}, {Stauffer}, {Megeath}, {Hora}, \&
  {Hartmann}}]{Rebull2006}
{Rebull}, L.~M., {Stauffer}, J.~R., {Megeath}, S.~T., {Hora}, J.~L., \&
  {Hartmann}, L. 2006, \apj, 646, 297, \dodoi{10.1086/504865}

\bibitem[{{Reiners} {et~al.}(2012){Reiners}, {Joshi}, \&
  {Goldman}}]{Reiners2012}
{Reiners}, A., {Joshi}, N., \& {Goldman}, B. 2012, \aj, 143, 93,
  \dodoi{10.1088/0004-6256/143/4/93}

\bibitem[{{Reinhold} \& {Hekker}(2020)}]{Reinhold2020}
{Reinhold}, T., \& {Hekker}, S. 2020, \aap, 635, A43,
  \dodoi{10.1051/0004-6361/201936887}

\bibitem[{{Ricker} {et~al.}(2015){Ricker}, {Winn}, {Vanderspek}, {Latham},
  {Bakos}, {Bean}, {Berta-Thompson}, {Brown}, {Buchhave}, {Butler}, {Butler},
  {Chaplin}, {Charbonneau}, {Christensen-Dalsgaard}, {Clampin}, {Deming},
  {Doty}, {De Lee}, {Dressing}, {Dunham}, {Endl}, {Fressin}, {Ge}, {Henning},
  {Holman}, {Howard}, {Ida}, {Jenkins}, {Jernigan}, {Johnson}, {Kaltenegger},
  {Kawai}, {Kjeldsen}, {Laughlin}, {Levine}, {Lin}, {Lissauer}, {MacQueen},
  {Marcy}, {McCullough}, {Morton}, {Narita}, {Paegert}, {Palle}, {Pepe},
  {Pepper}, {Quirrenbach}, {Rinehart}, {Sasselov}, {Sato}, {Seager},
  {Sozzetti}, {Stassun}, {Sullivan}, {Szentgyorgyi}, {Torres}, {Udry}, \&
  {Villasenor}}]{Ricker2015}
{Ricker}, G.~R., {Winn}, J.~N., {Vanderspek}, R., {et~al.} 2015, JATIS, 1,
  014003, \dodoi{10.1117/1.JATIS.1.1.014003}

\bibitem[{{Schmidt} {et~al.}(2015){Schmidt}, {Hawley}, {West}, {Bochanski},
  {Davenport}, {Ge}, \& {Schneider}}]{Schmidt2015}
{Schmidt}, S.~J., {Hawley}, S.~L., {West}, A.~A., {et~al.} 2015, \aj, 149, 158,
  \dodoi{10.1088/0004-6256/149/5/158}

\bibitem[{Shaw(2016)}]{GMOSPipeline}
Shaw, R.~A. 2016, {GMOS Data Reduction Cookbook, v. 1.2.2},  Tucson: National
  Optical Astronomy Observatory.
\newblock \url{http://ast.noao.edu/sites/default/files/GMOS_Cookbook/}

\bibitem[{{Skinner} {et~al.}(2017){Skinner}, {Morgan}, {West}, {L{\'e}pine}, \&
  {Thorstensen}}]{Skinner2017}
{Skinner}, J.~N., {Morgan}, D.~P., {West}, A.~A., {L{\'e}pine}, S., \&
  {Thorstensen}, J.~R. 2017, \aj, 154, 118, \dodoi{10.3847/1538-3881/aa83b5}

\bibitem[{{Skrutskie} {et~al.}(2006){Skrutskie}, {Cutri}, {Stiening},
  {Weinberg}, {Schneider}, {Carpenter}, {Beichman}, {Capps}, {Chester},
  {Elias}, {Huchra}, {Liebert}, {Lonsdale}, {Monet}, {Price}, {Seitzer},
  {Jarrett}, {Kirkpatrick}, {Gizis}, {Howard}, {Evans}, {Fowler}, {Fullmer},
  {Hurt}, {Light}, {Kopan}, {Marsh}, {McCallon}, {Tam}, {Van Dyk}, \&
  {Wheelock}}]{2mass}
{Skrutskie}, M.~F., {Cutri}, R.~M., {Stiening}, R., {et~al.} 2006, \aj, 131,
  1163, \dodoi{10.1086/498708}

\bibitem[{{Stassun} {et~al.}(2019){Stassun}, {Oelkers}, {Paegert}, {Torres},
  {Pepper}, {De Lee}, {Collins}, {Latham}, {Muirhead}, {Chittidi},
  {Rojas-Ayala}, {Fleming}, {Rose}, {Tenenbaum}, {Ting}, {Kane}, {Barclay},
  {Bean}, {Brassuer}, {Charbonneau}, {Ge}, {Lissauer}, {Mann}, {McLean},
  {Mullally}, {Narita}, {Plavchan}, {Ricker}, {Sasselov}, {Seager}, {Sharma},
  {Shiao}, {Sozzetti}, {Stello}, {Vanderspek}, {Wallace}, \&
  {Winn}}]{Stassun2019}
{Stassun}, K.~G., {Oelkers}, R.~J., {Paegert}, M., {et~al.} 2019, \aj, 158,
  138, \dodoi{10.3847/1538-3881/ab3467}

\bibitem[{{Stauffer} \& {Hartmann}(1986)}]{Stauffer1986}
{Stauffer}, J.~R., \& {Hartmann}, L.~W. 1986, \apjs, 61, 531,
  \dodoi{10.1086/191123}

\bibitem[{{Sunyaev} {et~al.}(2021){Sunyaev}, {Arefiev}, {Babyshkin},
  {Bogomolov}, {Borisov}, {Buntov}, {Brunner}, {Burenin}, {Churazov},
  {Coutinho}, {Eder}, {Eismont}, {Freyberg}, {Gilfanov}, {Gureyev}, {Hasinger},
  {Khabibullin}, {Kolmykov}, {Komovkin}, {Krivonos}, {Lapshov}, {Levin},
  {Lomakin}, {Lutovinov}, {Medvedev}, {Merloni}, {Mernik}, {Mikhailov},
  {Molodtsov}, {Mzhelsky}, {M{\"u}ller}, {Nandra}, {Nazarov}, {Pavlinsky},
  {Poghodin}, {Predehl}, {Robrade}, {Sazonov}, {Scheuerle}, {Shirshakov},
  {Tkachenko}, \& {Voron}}]{Sunyaev2021}
{Sunyaev}, R., {Arefiev}, V., {Babyshkin}, V., {et~al.} 2021, \aap, 656, A132,
  \dodoi{10.1051/0004-6361/202141179}

\bibitem[{{Tilley} {et~al.}(2019){Tilley}, {Segura}, {Meadows}, {Hawley}, \&
  {Davenport}}]{Tilley2019}
{Tilley}, M.~A., {Segura}, A., {Meadows}, V., {Hawley}, S., \& {Davenport}, J.
  2019, Astrobiology, 19, 64, \dodoi{10.1089/ast.2017.1794}

\bibitem[{{Torres-Robledo} {et~al.}(2020){Torres-Robledo}, {Brice{\~n}o},
  {Quint}, \& {Sanmartim}}]{GoodmanPipeline}
{Torres-Robledo}, S., {Brice{\~n}o}, C., {Quint}, B., \& {Sanmartim}, D. 2020,
  in Astronomical Society of the Pacific Conference Series, Vol. 522,
  Astronomical Data Analysis Software and Systems XXVII, ed. P.~{Ballester},
  J.~{Ibsen}, M.~{Solar}, \& K.~{Shortridge}, 533

\bibitem[{{VanderPlas}(2018)}]{VanderPlas2018}
{VanderPlas}, J.~T. 2018, \apjs, 236, 16, \dodoi{10.3847/1538-4365/aab766}

\bibitem[{{Vanderspek} {et~al.}(2019){Vanderspek}, {Huang}, {Vanderburg},
  {Ricker}, {Latham}, {Seager}, {Winn}, {Jenkins}, {Burt}, {Dittmann},
  {Newton}, {Quinn}, {Shporer}, {Charbonneau}, {Irwin}, {Ment}, {Winters},
  {Collins}, {Evans}, {Gan}, {Hart}, {Jensen}, {Kielkopf}, {Mao}, {Waalkes},
  {Bouchy}, {Marmier}, {Nielsen}, {Ottoni}, {Pepe}, {S{\'e}gransan}, {Udry},
  {Henry}, {Paredes}, {James}, {Hinojosa}, {Silverstein}, {Palle},
  {Berta-Thompson}, {Crossfield}, {Davies}, {Dragomir}, {Fausnaugh}, {Glidden},
  {Pepper}, {Morgan}, {Rose}, {Twicken}, {Villase{\~n}or}, {Yu}, {Bakos},
  {Bean}, {Buchhave}, {Christensen-Dalsgaard}, {Christiansen}, {Ciardi},
  {Clampin}, {De Lee}, {Deming}, {Doty}, {Jernigan}, {Kaltenegger}, {Lissauer},
  {McCullough}, {Narita}, {Paegert}, {Pal}, {Rinehart}, {Sasselov}, {Sato},
  {Sozzetti}, {Stassun}, \& {Torres}}]{Vanderspek2019}
{Vanderspek}, R., {Huang}, C.~X., {Vanderburg}, A., {et~al.} 2019, \apjl, 871,
  L24, \dodoi{10.3847/2041-8213/aafb7a}

\bibitem[{{Vida} {et~al.}(2017){Vida}, {K{\H{o}}v{\'a}ri}, {P{\'a}l},
  {Ol{\'a}h}, \& {Kriskovics}}]{Vida2017}
{Vida}, K., {K{\H{o}}v{\'a}ri}, Z., {P{\'a}l}, A., {Ol{\'a}h}, K., \&
  {Kriskovics}, L. 2017, \apj, 841, 124, \dodoi{10.3847/1538-4357/aa6f05}

\bibitem[{{West} \& {Basri}(2009)}]{West2009}
{West}, A.~A., \& {Basri}, G. 2009, \apj, 693, 1283,
  \dodoi{10.1088/0004-637X/693/2/1283}

\bibitem[{{West} {et~al.}(2004){West}, {Hawley}, {Walkowicz}, {Covey},
  {Silvestri}, {Raymond}, {Harris}, {Munn}, {McGehee}, {Ivezi{\'c}}, \&
  {Brinkmann}}]{West2004}
{West}, A.~A., {Hawley}, S.~L., {Walkowicz}, L.~M., {et~al.} 2004, \aj, 128,
  426, \dodoi{10.1086/421364}

\bibitem[{{West} {et~al.}(2011){West}, {Morgan}, {Bochanski}, {Andersen},
  {Bell}, {Kowalski}, {Davenport}, {Hawley}, {Schmidt}, {Bernat}, {Hilton},
  {Muirhead}, {Covey}, {Rojas-Ayala}, {Schlawin}, {Gooding}, {Schluns},
  {Dhital}, {Pineda}, \& {Jones}}]{West2011}
{West}, A.~A., {Morgan}, D.~P., {Bochanski}, J.~J., {et~al.} 2011, \aj, 141,
  97, \dodoi{10.1088/0004-6256/141/3/97}

\bibitem[{{Wright} {et~al.}(2011){Wright}, {Drake}, {Mamajek}, \&
  {Henry}}]{Wright2011}
{Wright}, N.~J., {Drake}, J.~J., {Mamajek}, E.~E., \& {Henry}, G.~W. 2011,
  \apj, 743, 48, \dodoi{10.1088/0004-637X/743/1/48}

\bibitem[{{Wright} {et~al.}(2018){Wright}, {Newton}, {Williams}, {Drake}, \&
  {Yadav}}]{Wright2018}
{Wright}, N.~J., {Newton}, E.~R., {Williams}, P. K.~G., {Drake}, J.~J., \&
  {Yadav}, R.~K. 2018, \mnras, 479, 2351, \dodoi{10.1093/mnras/sty1670}

\bibitem[{{York} {et~al.}(2000){York}, {Adelman}, {Anderson}, {Anderson},
  {Annis}, {Bahcall}, {Bakken}, {Barkhouser}, {Bastian}, {Berman}, {Boroski},
  {Bracker}, {Briegel}, {Briggs}, {Brinkmann}, {Brunner}, {Burles}, {Carey},
  {Carr}, {Castander}, {Chen}, {Colestock}, {Connolly}, {Crocker}, {Csabai},
  {Czarapata}, {Davis}, {Doi}, {Dombeck}, {Eisenstein}, {Ellman}, {Elms},
  {Evans}, {Fan}, {Federwitz}, {Fiscelli}, {Friedman}, {Frieman}, {Fukugita},
  {Gillespie}, {Gunn}, {Gurbani}, {de Haas}, {Haldeman}, {Harris}, {Hayes},
  {Heckman}, {Hennessy}, {Hindsley}, {Holm}, {Holmgren}, {Huang}, {Hull},
  {Husby}, {Ichikawa}, {Ichikawa}, {Ivezi{\'c}}, {Kent}, {Kim}, {Kinney},
  {Klaene}, {Kleinman}, {Kleinman}, {Knapp}, {Korienek}, {Kron}, {Kunszt},
  {Lamb}, {Lee}, {Leger}, {Limmongkol}, {Lindenmeyer}, {Long}, {Loomis},
  {Loveday}, {Lucinio}, {Lupton}, {MacKinnon}, {Mannery}, {Mantsch}, {Margon},
  {McGehee}, {McKay}, {Meiksin}, {Merelli}, {Monet}, {Munn}, {Narayanan},
  {Nash}, {Neilsen}, {Neswold}, {Newberg}, {Nichol}, {Nicinski}, {Nonino},
  {Okada}, {Okamura}, {Ostriker}, {Owen}, {Pauls}, {Peoples}, {Peterson},
  {Petravick}, {Pier}, {Pope}, {Pordes}, {Prosapio}, {Rechenmacher}, {Quinn},
  {Richards}, {Richmond}, {Rivetta}, {Rockosi}, {Ruthmansdorfer}, {Sandford},
  {Schlegel}, {Schneider}, {Sekiguchi}, {Sergey}, {Shimasaku}, {Siegmund},
  {Smee}, {Smith}, {Snedden}, {Stone}, {Stoughton}, {Strauss}, {Stubbs},
  {SubbaRao}, {Szalay}, {Szapudi}, {Szokoly}, {Thakar}, {Tremonti}, {Tucker},
  {Uomoto}, {Vanden Berk}, {Vogeley}, {Waddell}, {Wang}, {Watanabe},
  {Weinberg}, {Yanny}, {Yasuda}, \& {SDSS Collaboration}}]{york00}
{York}, D.~G., {Adelman}, J., {Anderson}, John~E., J., {et~al.} 2000, \aj, 120,
  1579, \dodoi{10.1086/301513}

\bibitem[{{Zechmeister} \& {K{\"u}rster}(2009)}]{Zechmeister2009}
{Zechmeister}, M., \& {K{\"u}rster}, M. 2009, \aap, 496, 577,
  \dodoi{10.1051/0004-6361:200811296}

\bibitem[{{Ziegler} {et~al.}(2018){Ziegler}, {Law}, {Baranec}, {Morton},
  {Riddle}, {De Lee}, {Huber}, {Mahadevan}, \& {Pepper}}]{Ziegler2018}
{Ziegler}, C., {Law}, N.~M., {Baranec}, C., {et~al.} 2018, \aj, 156, 259,
  \dodoi{10.3847/1538-3881/aad80a}

\end{thebibliography}
\bibliographystyle{aasjournal}

\appendix\label{sec:appendix}
\vspace{.5in}

\begin{figure}
    \centering
    \includegraphics[scale=1.1]{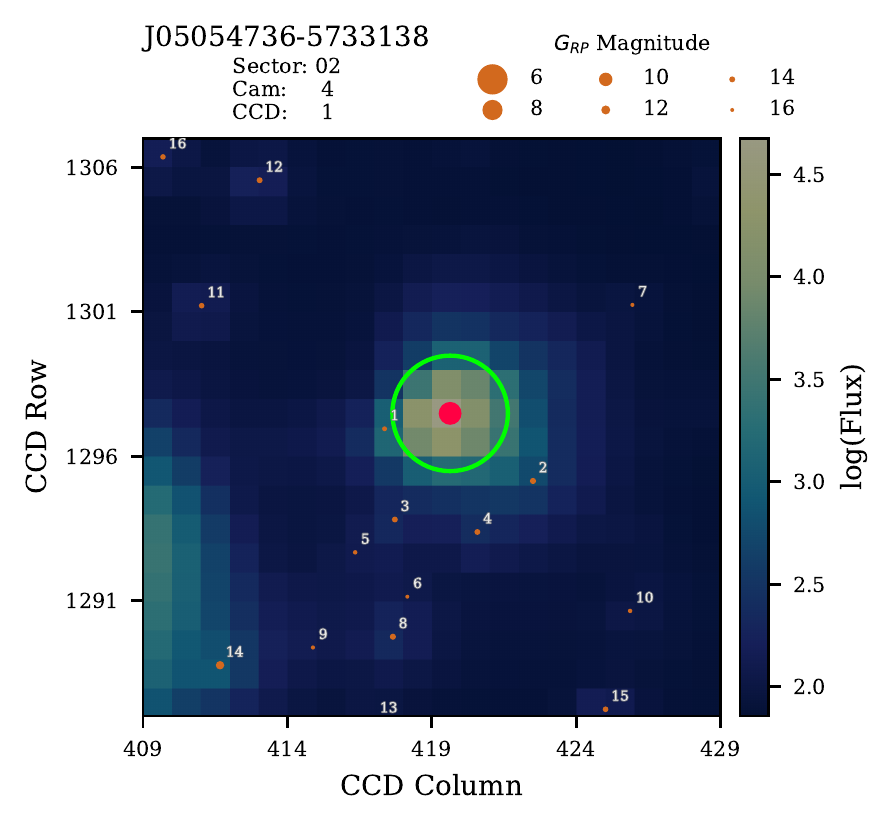} \caption{The 21 $\times$ 21 pixel cutout of the TESS full-frame image for 2MASS J$05054736-5733138$ (red dot). The green circle has a two-pixel radius and corresponds to the aperture used to extract the light curve. The color scale represents the flux intensity. The numbered points show the nearby Gaia sources; their size is proportional to their $G_{\rm RP}$ mag. Information such as the TESS sector, CCD number, and camera are also shown. The complete figure set (12 images) is available in the online journal.}
    \label{fig:tpexample}
\end{figure}

\begin{figure}
    \centering
    \includegraphics[scale=1]{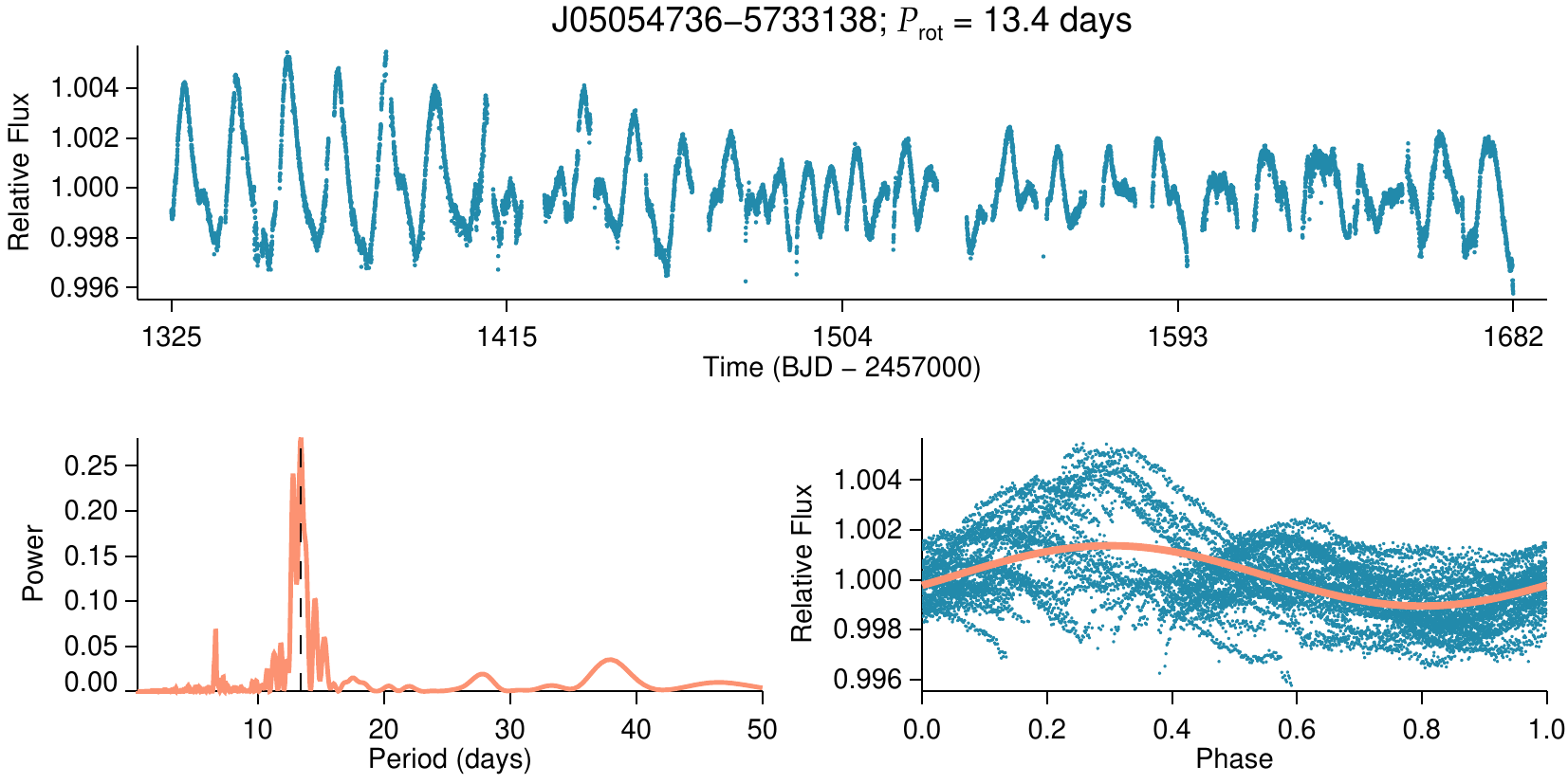}
    \caption{{\it Top}: Normalized TESS light curve. {\it Bottom left}: The GLS periodogram analysis shows a peak at \prot = 13.4 days (vertical dash line). {\it Bottom right}: Phase-folded light curve to this \prot, which visually helps to confirm the period determination. The complete figure set (12 images) is available in the online journal.
    }
    \label{fig:lcexample}
\end{figure}

\end{document}